\documentclass[twocolumn]{aastex62}

\graphicspath{{./}{figures/}}

\received{July 30, 2019}
\revised{November 11, 2019}
\accepted{November 11, 2019}
\submitjournal{ApJS}

\shorttitle{New hard-TeV blazars with MAGIC}
\shortauthors{MAGIC Collaboration}

\usepackage{natbib}
\usepackage{multirow}
\usepackage{threeparttable}
\usepackage{subfigure}
\usepackage{gensymb}
\usepackage{soul}
\usepackage{footmisc}
\NewPageAfterKeywords

\begin{document}

\title{New hard-TeV extreme blazars detected with the MAGIC telescopes}

\collaboration{(MAGIC Collaboration)}

\author[0000-0001-8307-2007]{V.~A.~Acciari} 
\affiliation{Inst. de Astrof\'isica de Canarias, E-38200 La Laguna, and Universidad de La Laguna, Dpto. Astrof\'isica, E-38206 La Laguna, Tenerife, Spain}

\author[0000-0002-5613-7693]{S.~Ansoldi} 
\affiliation{Universit\`a di Udine, and INFN Trieste, I-33100 Udine, Italy}
\affiliation{Japanese MAGIC Consortium: ICRR, The University of Tokyo, 277-8582 Chiba, Japan; Department of Physics, Kyoto University, 606-8502 Kyoto, Japan; Tokai University, 259-1292 Kanagawa, Japan; RIKEN, 351-0198 Saitama, Japan}

\author[0000-0002-5037-9034]{L.~A.~Antonelli} 
\affiliation{National Institute for Astrophysics (INAF), I-00136 Rome, Italy}

\author{A.~Arbet Engels} 
\affiliation{ETH Zurich, CH-8093 Zurich, Switzerland}

\author{K.~Asano} 
\affiliation{Japanese MAGIC Consortium: ICRR, The University of Tokyo, 277-8582 Chiba, Japan; Department of Physics, Kyoto University, 606-8502 Kyoto, Japan; Tokai University, 259-1292 Kanagawa, Japan; RIKEN, 351-0198 Saitama, Japan}

\author{D.~Baack} 
\affiliation{Technische Universit\"at Dortmund, D-44221 Dortmund, Germany}

\author[0000-0002-1444-5604]{A.~Babi\'c} 
\affiliation{Croatian Consortium: University of Rijeka, Department of Physics, 51000 Rijeka; University of Split - FESB, 21000 Split; University of Zagreb - FER, 10000 Zagreb; University of Osijek, 31000 Osijek; Rudjer Boskovic Institute, 10000 Zagreb, Croatia}

\author[0000-0002-8008-2485]{B.~Banerjee} 
\affiliation{Saha Institute of Nuclear Physics, HBNI, 1/AF Bidhannagar, Salt Lake, Sector-1, Kolkata 700064, India}

\author[0000-0001-7909-588X]{U.~Barres de Almeida} 
\affiliation{Centro Brasileiro de Pesquisas F\'isicas (CBPF), 22290-180 URCA, Rio de Janeiro (RJ), Brasil}

\author[0000-0002-0965-0259]{J.~A.~Barrio} 
\affiliation{IPARCOS Institute and EMFTEL Department, Universidad Complutense de Madrid, E-28040 Madrid, Spain}

% ...



\begin{abstract}
Extreme high-frequency peaked BL Lac objects (EHBLs) are blazars which exhibit extremely energetic synchrotron emission. They also feature non-thermal gamma-ray emission whose peak lies in the very high-energy (VHE, $E>100$\,GeV) range, and in some sources exceeds 1\,TeV: this is the case of hard-TeV EHBLs such as 1ES~0229+200. With the aim of increasing the EHBL population, ten targets were observed with the MAGIC telescopes from 2010 to 2017, for a total of  262\,h of good quality data.  The data were complemented by  coordinated {\it Swift} observations. The X-ray data analysis confirms that all the sources but two are EHBLs. The sources show only a modest variability and a harder-when-brighter behavior, typical for this class of objects. At VHE gamma rays, three new sources were detected and a hint of signal was found for another new source. In each case the intrinsic spectrum is compatible with the hypothesis of a hard-TeV nature of these EHBLs. The broadband spectral energy distributions (SEDs) of all sources are built and modeled in the framework of a single-zone purely leptonic model. The VHE gamma-ray detected sources  were also interpreted with a spine-layer and a proton synchrotron models. The three models provide a good description of the SEDs. However, the resulting  parameters differ substantially  in the three scenarios, in particular the magnetization parameter. This work presents a first mini-catalog of VHE gamma-ray and multi-wavelength observations of EHBLs.
\end{abstract}

\keywords{Catalogs	 - Active galaxies - galaxy jets - BL Lacertae objects - Gamma-ray sources - Non-thermal radiation sources}

\section{\label{sec:intro}Introduction} 

Giant elliptical galaxies may host in their center a super-massive black hole ($\sim$10$^{9}$ M$_{\odot}$) which accretes material through a disc and, in 1 up to 15\% of the cases \citep{2017AARv..25....2P}, features two narrow jets of ultra-relativistic particles extending well outside the galaxy. These objects are known as jetted active galactic nuclei (jetted-AGNs; \citealt{Urry95,2016AARv..24...13P}). The spectra observed from jetted-AGNs is strongly dependent on the viewing angle of the jet with respect to the observer. This difference is also at the base of their classification. Radio galaxies are jetted-AGNs with the jets seen from large angles. The two extended jets are particularly bright in radio and gamma rays. Blazars are instead jetted-AGNs seen at small angles, and their spectra is fully dominated by the jet emission which is largely enhanced due to relativistic effects. They can be subdivided into flat spectrum radio quasars (FSRQs) and BL Lac objects (BL Lacs) depending on the equivalent widths of emission lines in the optical spectrum \citep{1991ApJS...76..813S,1991ApJ...374..431S}. \citet{ghisellini09} suggested that the division between these two classes is due to the different accretion regime, with FSRQs showing an accretion rate above $10^{-2}$ of the Eddington rate. The spectral energy distribution (SED) emitted by blazars is characterized by two broad humps \citep{ghisellini17}: a low-frequency (from $\sim$10$^{12}$ to $10^{18}$\,Hz and above), and a high-frequency peak (above $10^{21}$\,Hz). The first peak is due to synchrotron radiation emitted by ultra-relativistic electrons. The  second peak is most likely due to inverse Compton (IC) emission and is possibly accompanied by an additional hadronic component whose relevance is still largely debated \citep{2013ApJ...768...54B}. The location of the first peak is on average at quite low frequencies for FSRQs, and drives the division of BL Lacs into the sub-categories LBL, IBL, and HBL (low-, intermediate-, and high-frequency-peaked BL Lacs, respectively).  \citet{Fossati98} found evidence of an empirical sequence connecting the blazar classes with their bolometric luminosity, that is, low-energy-peaked objects such as FSRQs display a higher luminosity than high-energy-peaked ones, i.e., HBLs, and form the so-called blazar sequence. In addition, the luminosity ratio between the high and low energy component increases with bolometric luminosity. According to \citet{ghisellini98}, this anti-correlation between the peak position of the synchrotron emission and the bolometric luminosity can be explained by effective cooling effects. Effective cooling is more efficient for FSRQs due to the strong radiation fields within the broad line region (BLR). This leads to a lower Lorentz factor at the break of the electron distribution, which determines the location of both the synchrotron and the Compton peaks, and therefore largely determines the shape of the SED.

The other important parameters characterizing the SED of blazars are the ratio of the Compton-to-synchrotron powers, i.e., the Compton dominance, the power injected in the form of electrons, and the power in the external photon component. Since external radiation fields are present in FSRQs, this latter component contributes to effective cooling. Based on blazars with known redshift that have been detected by the Large Area Telescope (LAT) on board the \textit{Fermi Gamma-ray Space Telescope}, \citet{ghisellini17} revise the blazar sequence. The authors report to find a sequence with the same general properties of the original one. In addition, when considering BL Lacs and FSRQs separately, they find that FSRQs form a sequence in Compton dominance and in the X-ray spectral index. However, they do not become redder when being more luminous, while BL Lacs do.

In this context, \citet{2001AA...371..512C} found evidence of objects with the synchrotron peak frequency exceeding the soft X-ray band, defined as extreme high-frequency-peaked blazars (EHBLs,  peak above 10$^{17}$\,Hz, see also \citealt{2010ApJ...716...30A}). According to the blazar sequence, these objects are expected to be very faint, being at the upper edge of the peak frequency location. However, several observation campaigns in multi-bands carried out on blazars have found evidences of a number of relatively bright EHBLs (e.g., 1ES~1426+428, \citealt{2001AA...371..512C}) as well as two blazars classified as HBLs that show during flaring states EHBLs behavior (e.g., Mrk~501 and 1ES~2234+514, \citealt{Ghisellini99}), which are somehow in contradiction with the blazar sequence (e.g., \citealt{Padovani07,Giommi11,Kaur18}).

In the last decade, the very good performances of running Imaging Atmospheric Cherenkov Telescopes (IACTs; namely H.E.S.S., MAGIC, and VERITAS) opened the possibility of observing this intriguing class of objects at very-high energies (VHE, $E > 100$\,GeV). 
VHE gamma-ray observations are distance limited, due to the interaction of VHE photons with the extragalactic background light (EBL) which causes a suppression of the gamma-ray flux. This suppression increases with the distance of the source and with the energy of VHE photons: for nearby sources ($z < 0.05$) it is effective only above few TeV, but for relatively distant sources  (z $>$ 0.5) it is effective already at few hundred GeV. At $z\sim1.0$, 100 GeV photons are already strongly absorbed (e.g., \citealt{2008AA...487..837F}). The current catalog of extragalactic sources detected at VHE by IACTs (TeVCat\footnote{\href{tevcat.uchicago.edu}{tevcat.uchicago.edu}}) counts $\sim$80 objects. The large majority are HBLs with a high-energy SED peak located typically at or above 100\,GeV. Out of these sources, there are 14 sources with published spectra cataloged as EHBLs \citep{foffano19,magic19}.

There are seven objects detected at TeV energies and classified in \citet{2018MNRAS.477.4257C} and \citet{magic19} as hard-TeV blazars, with a second SED bump peaking above 1\,TeV. This translates in a VHE power-law spectral index in the $100$\,GeV--$1$\,TeV range smaller than 2. Other 7 objects are EHBLs with a  softer TeV spectra \citep{foffano19}. Interestingly, at least other two sources (Mrk~501, \citealt{1998ApJ...492L..17P,2018AA...620A.181A}; and 1ES~1959+650, \citealt{2018AA...620A.181A}) have shown  EHBL behavior (and hard TeV spectra) during flaring states.  As discussed in \citet{foffano19}, these different behaviors at VHE gamma rays might be characterizing different sub-classes within the EHBL class. Among TeV-detected EHBLs, 1ES~0229+200 has the highest high-energy peak frequency.


From the phenomenological and theoretical point of view, the spectral characteristics of hard-TeV EHBLs make these sources extremely interesting objects to be studied in further detail. The  prototypical  hard-TeV EHBL is 1ES~0229+200, located at a moderate redshift of 0.14 \citep{aharonian07,tavecchio09}. The synchrotron peak of 1ES~0229+200 was sampled in great detail in a multi-wavelength campaign carried out in 2010 including optical, UV and X-ray data which firmly characterized the synchrotron emission of this object \citep{kaufmann11,aliu14}. The high X-ray/UV flux ratios that were observed indicate a remarkably hard synchrotron spectrum, which could be a hint for the presence of a low-energy cutoff of the electron spectrum \citep{kaufmann11}. Once corrected for EBL absorption, the VHE gamma-ray spectrum indicates a flux that is steadily increasing with energy, suggesting that in this object the high-energy bump of the SED exceeds few TeV \citep{aharonian07}. 

\begin{table*}
\centering
\caption{\label{tab:src_table}Sample of EHBLs observed with MAGIC.Columns from \textit{left} to \textit{right}: source name, equatorial (RA and DEC) and Galactic coordinates (l and b), redshift ($z$), equivalent Galactic hydrogen column density reported by \citet{2005AA...440..775K}, synchrotron peak frequency reported by \citet{Chang17} ($\mathrm{log}$($\nu_{\rm{peak}}$)), criteria adopted for the selection (see text for details).
1ES~0229+200 reported in the last line is the prototype of EHBL sources and is considered in our work as reference source.}
\begin{tabular}{lcccccccc}
\hline
\multirow{2}{*}{Source} & RA (J2000)   & DEC (J2000) & l     & b       & \multirow{2}{*}{$z$} & $N_{H}$& $\mathrm{log}$($\nu_{\rm{peak}}$) & Selection \\
& [$\degree$] & [$\degree$]  &  [$\degree$]     &   [$\degree$]     & &  $\times 10^{21}$\,[cm$^{-2}$]  &   [Hz]          & Criteria  \\
\hline
TXS~0210+515     & 33.57   &  51.75  & 135.74 & -9.05  & 0.049$^\mathrm{1}$     & 1.440 & 17.3 & i, ii, iv, v     \\
TXS~0637-128     & 100.03  & -12.89  & 223.21  &-8.31  & 0.136$^\mathrm{2}$  & 2.990 & 17.4 & ii, v       \\
BZB~J0809+3455   & 122.41  &  34.93  & 186.48 & 30.35  & 0.082$^\mathrm{3}$     & 0.432 & 16.6 & i, ii, iv, v      \\
RBS~0723         & 131.80  &  11.56  & 215.46 & 30.89  & 0.198$^\mathrm{3}$     & 0.317 & 17.8 & i, ii, iii, v   \\
1ES~0927+500     & 142.66  &  49.84  & 168.14 & 45.71  & 0.187$^\mathrm{3}$     & 0.138 & 17.5 & iii, v         \\
RBS~0921         & 164.03  &  2.87   & 249.28 & 53.28  & 0.236$^\mathrm{3}$     & 0.382 & 17.9 & iii         \\ 
1ES~1426+428     & 217.14  &  42.70  & 77.48  & 64.90  & 0.129$^\mathrm{3}$     & 0.113 & 18.1 & i, ii, v      \\ 
1ES~2037+521     & 309.85  &  52.33  & 89.69  & 6.55   & 0.053$^\mathrm{1}$     & 4.360 & N.A. & i, ii, iv, v      \\
RGB~J2042+244    & 310.53  &  24.45  & 67.77  & -10.80 & 0.104$^\mathrm{4}$
& 1.010 & 17.5 & ii, v         \\
RGB~J2313+147    & 348.49  &  14.74  & 90.5   & -41.91 & 0.163$^\mathrm{5}$    & 0.514 & 17.7 & ii, v         \\
\hline
1ES~0229+200     & 38.20   &  20.29  & 152.94 & -36.61 & 0.140$^\mathrm{1}$      & 0.792 & 18.5 & -         \\
\hline
\end{tabular}
\begin{tablenotes}
\item {1: \citealt{mao11}; 2: private communication with S. Paiano; 3: \citealt{ahn12}; 4: \citealt{shaw13}; 5: \citealt{sowards05}}
\end{tablenotes}
\end{table*}

Since the detection of its peculiar TeV spectrum, 1ES~0229+200 became of fundamental importance for the EBL science case and for constraining the intergalactic magnetic field (IGMF). Due to the extreme hardness of the intrinsic spectrum which does not show any curvature at VHE up to 10\,TeV, 1ES~0229+200 yields the necessary TeV photons to study a wider range of the EBL spectrum up to the, yet less constrained, far infrared band \citep{aharonian07}. In the cosmological context a high intrinsic energy up to 10 TeV is a requisite to derive  limits on the IGMF \citep{murase12}. In fact, the photons emitted {above 1 TeV} from distant EHBLs lead to electromagnetic cascades sensitive to the magnetic field in the intergalactic medium. The IGMF leaves its imprint in the reprocessed gamma-rays, resulting in an excess in the GeV {energy} range that can be measured with instruments like {\it Fermi}/LAT \citep{vovk12}.

The number of relevant studies carried out on 1ES~0229+200 justifies and supports the need for deep observations on other objects with similar properties. These studies, in fact, suffer from the very limited sample of hard-TeV EHBLs known both in X-rays and VHE gamma rays. Considering the extreme properties of their peak components, the investigation of their X-ray and VHE gamma-ray emission {is the main goal of the present study. Moreover, it is the first and most important building block to address all of the scientific outcomes briefly introduced above.}

It is important to notice that in the high-energy gamma-ray band (HE; 100\,MeV $<$ E $<$ 100\,GeV) faint hard-TeV EHBLs are objects {that are very difficult to detect}. This is due to a combination of the average low-luminosity characteristics for this kind of objects and the high-energy peak of the SED located around or above 1 TeV. {For example, the {\it Fermi}-LAT reports a significant detection of 1ES~0229+200 only after 4 years of exposure time \citep{Acero15,vovk12} and despite the hard VHE spectrum it is not present in the the Second Catalog of Hard \textit{Fermi}-LAT Sources, 2FHL \citep{Ackermann16}.} 

The paper is structured as follows: in Section~\ref{sec:sources}, a short description of the criteria adopted for the source selection is given followed by a list of the ten targets {of this study}. Sections~\ref{sec:MAGIC},~\ref{sec:HE}, and~\ref{sec:X-ray_prop} report the results of the MAGIC, {\it Fermi}-LAT, {\it Swift}-XRT and {\it NuSTAR} data analysis, respectively. Section~\ref{sec:X-ray_prop} includes a study of the X-ray temporal properties of the sample. The observational properties of the sources in other bands are briefly outlined in Section~\ref{sec:OtherBands}.  The multi-wavelength SED data and models are reported and discussed in Section~\ref{sec:modelling}. Finally, Section~\ref{sec:discuss} includes a final discussion and a summary of the main results of the paper. The details of the data analyses in the various bands as well as those of the modeling are reported in the Appendices~\ref{app:magic} to~\ref{app:SED0921}.

\section{\label{sec:sources}Source Selection}

Regarding the selection of EHBL targets for the observation with the MAGIC telescopes, different approaches have been attempted.  Such an approach facilitated the chances of detection and take{s} the updated catalogs into consideration. The general criteria adopted are based on the X-ray spectral behavior, the soft HE gamma-ray spectral {behavior}, and the X-ray-to-radio flux ratio. 

The first criterion (i) relies on the fact that EHBLs are by definition expected to exhibit the synchrotron peak above  $10^{17}$\,Hz. 
Therefore, candidates with a hard spectral index ($\Gamma \leq 2$) in the soft X-ray band covered by {\it Swift}-XRT were targeted. Additionally, the tail of the synchrotron emission could be also detected at hard X-rays by {\it Swift}/BAT and {\it NuSTAR}. 

The second criterion (ii) adopted for the selection is  related to the properties of the HE  gamma-ray emission of each source extracted from the following LAT {catalogs}: the 1FHL, the First \textit{Fermi}-LAT Catalog of Sources above 10 GeV \citep{Ackermann13}, the 2FHL, the Second Catalog of Hard \textit{Fermi}-LAT Sources \citep{Ackermann16}, and the 3FGL, \textit{Fermi}-LAT 4-year Point Source Catalog \citep{Acero15}.  The second peak of the SED of EHBLs might be difficult to measure below a hundred GeV, especially {when it is located above 1\,TeV}. This is {for example} the case of 1ES~0229+200, whose second SED peak was constrained above 10 TeV by H.E.S.S. and VERITAS observations. On the other hand, a possible detection, even if marginal, of gamma rays in the HE gamma-ray range enhances significantly the chance of the detectability with MAGIC, and makes the extrapolation to the VHE possible. For this reason, the gamma-ray emission properties as reported in the LAT catalogs, {when available}, {have} been considered for the selection of new candidates.

In recent MAGIC observation campaigns the list of EHBL candidates proposed in \citet{Bonnoli15}, where the authors propose new candidates according to the high X-ray-to-radio flux ratio, was considered. This was the third selection criterion  (iii).

\citet{2017AA...608A..68F} proposed a list of 53 promising TeV BL Lac candidates based on the multi-wavelength luminosity correlations derived for the sample of TeV-detected BL Lac objects. As the forth criterion (iv) we selected the best candidates {whose} X-rays and HE gamma-ray properties follow criteria (i) and (ii).

Finally, {low-redshift ($<$0.2) sources were favored} in the selection {as criterion (v)}, ensuring a relatively small effect on the VHE spectra due to EBL absorption, at least below the TeV range.

The sources whose MAGIC spectrum is already published, e.g. 1ES~1741+196 and the recently detected 2WHSP~J073326.7+515354 \citep{2017MNRAS.468.1534A,magic19}, or collected after 2017 have been excluded from the sample.

The final list of objects observed with the MAGIC telescopes is summarized in Table~\ref{tab:src_table}. The equatorial and Galactic coordinates of the sources are listed together with the redshift, Equivalent Galactic hydrogen column density reported by \citet{2005AA...440..775K}, and the synchrotron peak frequency as reported in the 2WHSP (Second Wise HSP catalog; \citealt{Chang17}), when available. The last column summarizes the criteria used for the selection. 

The sample includes the archetypal EHBL source 1ES~0229+200, which has been deeply observed by MAGIC between 2013 and 2017 and is added as a reference source \citep[MAGIC Coll. in prep.]{2019MNRAS.486.4233A}. All the {considered sources} have not been detected by IACTs {except for} 1ES~1426+428, which was first discovered as a TeV emitter by Whipple \citep{2002AA...384L..23A} and recently detected with the VERITAS telescopes \citep{2017ApJ...835..288A}. 

All the selected sources show a hard spectral index in the X-ray band and, {except for} RBS~0921, are all listed in the 3FGL catalog. Moreover, all the sources selected are  present in the 2WHSP of high-synchrotron-peaked blazars except for 1ES~2037+521, whose very bright host galaxy is probably the cause of exclusion from the 2WHSP selection.

\section{\label{sec:MAGIC}MAGIC Results}

Ten targets were observed with the MAGIC telescopes starting from 2010. A total of 262\,h of good quality data were collected and analyzed. Table~\ref{tab:obs_table_magic} summarizes  the general information of MAGIC observations. 
A fraction of the data was collected during moderate moon time, which explains the relatively high energy threshold reported. {The details of the analysis of data taken with the MAGIC telescopes are reported in Appendix~\ref{app:magic}.} 

\begin{table*}
\centering
\caption{\label{tab:obs_table_magic} Results of the signal search and integral flux analysis of the MAGIC data for the ten EHBLs considered in this study. The results for 1ES~0229+200 are also reported for comparison, bottom row. Columns from \textit{left} to \textit{right}: source name, year(s) of observation, effective exposure time after quality cuts, significance of the signal in $\sigma$,  assumed energy threshold for integral flux calculation, flux measured above the energy threshold, VHE gamma-ray luminosity over 200 GeV, and the source detection status at VHE gamma rays (Y: detected, N: not detected, and H: hint of signal). In case of non-detection (see Section\,\ref{sec:MAGIC} for details), an integral-flux upper limit is reported instead, assuming a simple power-law spectrum of spectral index $\Gamma$ (see Equation~\ref{eq:spec} and the text for further details).}
\begin{tabular}{lccccccc}
\hline
\multirow{2}{*}{Source}    & \multirow{2}{*}{Observation periods}             & Time  & Significance & $E_{\mathrm{th}}$ & Flux$_{\geq E_{\mathrm{th}}}$ & L$_{\geq 200\mathrm{GeV}}$ & \multirow{2}{*}{VHE?} \\
&                      & [h] & [$\sigma$]      & [GeV]       & $\times 10^{-12}$[cm$^{-2}$s$^{-1}$] & $\times 10^{43}$[erg\,s$^{-1}$] &\\
                  	
\hline

TXS~0210+515   & 2015, 2016, 2017 & 28.6  & 5.9  & 200 & 1.6 $\pm$ 0.5  & 0.6  $\pm$ 0.2 & Y\\
TXS~0637-128   & 2017     		  & 12.8  & 1.7  & 300 & $<$8.9$^\star$ & $<$50.9 & N\\
BZB~J0809+3455 & 2015             & 21.8  & 0.4  & 150 & $<$3.7$^\star$ & $<$3.0  & N\\
RBS~0723       & 2013, 2014       & 45.3  & 5.4  & 200 & 2.6 $\pm$ 0.5  & 24.8 $\pm$ 4.8 & Y\\
1ES~0927+500   & 2012, 2013       & 26.2  & 1.2  & 150 & $<$5.1$^\star$ & $<$24.2 & N\\
RBS~0921       & 2016             & 13.9  & -0.4 & 150 & $<$8.6$^\star$ & $<$68.5 & N\\  
\multirow{3}{*}{1ES~1426+428 \Bigg\{}&2010 & 6.51 & 2.1  & 200 & $<$9.3$^\dagger$ & $<$27.7 & N \\
               & 2012             & 8.7   & 6.0  & 200 & 6.1 $\pm$ 1.1  & 18.4 $\pm$ 3.4 & Y\\
               & 2013             & 5.9   & 1.8  & 200 & $<$5.1$^\dagger$ & $<$14.2 & N\\ 
1ES~2037+521   & 2016             & 28.1  & 7.5  & 300 & 1.8 $\pm$ 0.4  & 1.3  $\pm$ 0.3 & Y\\
RGB~J2042+244  & 2015             & 52.5  & 3.7  & 200 & 1.9 $\pm$ 0.5  & 3.4  $\pm$ 0.8 & H\\
RGB~J2313+147  & 2015             & 11.5  & -0.9 & 200 & $<$1.5$^\star$ & $<$7.0  & N\\
\hline
1ES~0229+200  &  2013--2017       & 117.5 & 9.0  & 200 & 2.1 $\pm$ 0.3  & 7.6  $\pm$ 1.1 & Y\\
\hline
\end{tabular}
\begin{tablenotes}
\item$\star$ Flux upper limit is calculated by assuming the observed photon index $\Gamma_{\text{obs}}=2.0$.
\item$\dagger$ Flux upper limit is calculated by assuming the observed photon index $\Gamma_{\text{obs}}=2.6$ derived from 2012 observations.
\end{tablenotes}
\end{table*}

For comparison, the results of the analysis of 117.46\,h of 1ES~0229+200 data collected with the MAGIC telescopes between 2013 and 2017 (MAGIC Coll. in prep.) are also reported. 
The significance of the signal from this source is  $9 \sigma$: although the second SED peak lies in the TeV range its overall luminosity is low, as predicted by the blazar sequence, and therefore it does not reach a very high significance despite the long exposures. 

\subsection{\label{subsec:MAGIC_signal}Signal Search and integral flux analysis}
For the signal search, the $\theta^2$ method explained in  Appendix~\ref{app:magic} was adopted. The significance of the gamma-ray signal, estimated with formula [17] of \citet{LiMa83}, is reported in the fourth column of Table~\ref{tab:obs_table_magic}. 

The analysis revealed firm VHE gamma-ray detection of three new sources, namely TXS~0210+515, RBS~0723, and 1ES~2037+521, {and a hint of signal from RGB~J2042+244. In addition, a firm detection of the known TeV emitter 1ES~1426+428 was found in the 2012 dataset.} A dedicated time-resolved analysis was performed on each source. In particular a {possible daily-, monthly- and yearly-scale} variability was checked {and n}o hint of variability in the {analyzed} sample was detected. For 1ES~1426+428, a yearly-scale analysis resulted in a significant signal detection only from the 2012 dataset (see Fig.~\ref{Fig:1426_LC} in Appendix~\ref{app:magic}). However, with the data collected the constant-flux hypothesis cannot  be excluded ($\chi^2/\mathrm{d.o.f}=8.439/2$;  d.o.f. = degrees of freedom). 1ES~1426+428 is the only source of the sample previously detected by IACTs \citep{cat00,veritas01,whipple02,hegra02,delaCalle03,hegra03,veritas11,gt4812}. A comparison of the integral flux and of the observed spectra  can be found in Appendix\,\ref{app:magic}.

\citet{archambault16} reports VHE gamma-ray flux upper limits {obtained with the VERITAS array} for four sources in our sample. They are TXS~0210+515, BZB~J0809+3455, 1ES~0927+500, and RBS~0921. Among these sources, the VHE gamma-ray flux of TXS~0210+515 measured during MAGIC campaign is in agreement with the upper limit reported by VERITAS, which lies above MAGIC measurement. In the other three cases, MAGIC observations led to a better constraint of VHE gamma-ray flux when comparing the reported upper limits by VERITAS. This reflects the deeper exposures adopted by the MAGIC Collaboration. Regarding the variability, it must be underlined that all the sources considered are faint TeV emitters and a possible moderate variability of the signal could be undetectable due to the instrument's sensitivity limit.

\subsection{\label{subsec:MAGIC_spectra}Spectral Analysis}

\begin{figure}
\centering
\includegraphics[width=0.47\textwidth]{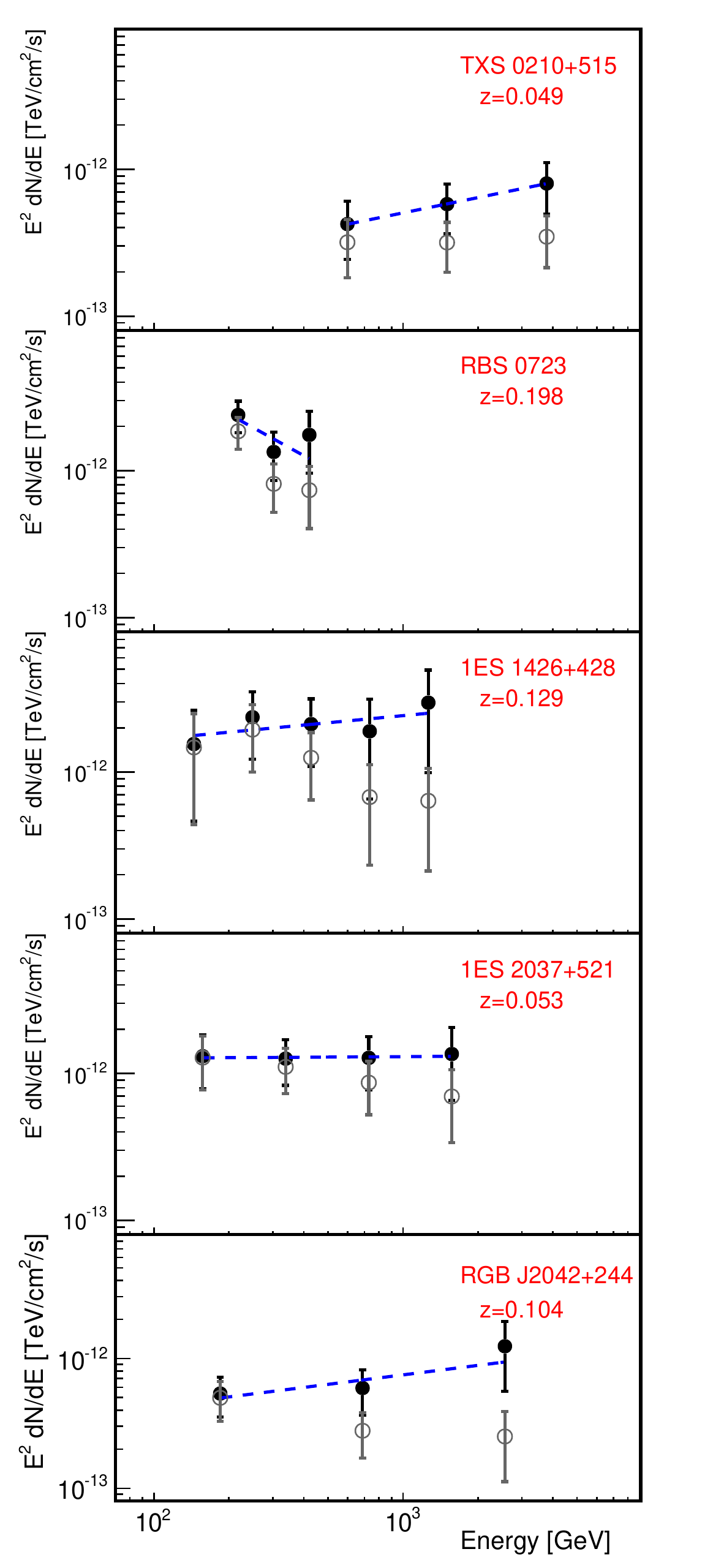}
\caption{\label{fig:MAGICSpectra}Spectral energy distributions of the four extreme blazars detected with the MAGIC telescopes and for the hint-of-signal source in E$^{2}$ dN/dE representation: observed data (open {gray} markers) and EBL-corrected data (filled black markers) using the \citet{2008AA...487..837F} model. The dashed lines represent the power-law fit to the EBL-corrected data.}
\end{figure}

The observed spectra of the {three new} sources detected with MAGIC, 1ES~1426+428, and for the hint-of-signal source are displayed in E$^{2}dN/dE$ representation in Figure~\ref{fig:MAGICSpectra} as open {gray} markers.

All the spectra are characterized by only three to five spectral points that are affected by large uncertainties due to the relatively faint signals.  
Interestingly, all the sources except the most distant one, that is RBS 0723, display data points above 1\,TeV, which excludes severe cutoff below this energy as expected for this class of sources, {in particular the hard-TeV ones}.

The spectra have been fitted with a simple power law of the form 
\begin{equation}\label{eq:spec}
    \frac{dN}{dE} = F_0 \cdot \left( 
    \frac{E}{E_{\rm{dec}}}\right)^{-\Gamma},
\end{equation}
with $F_0$ and $\Gamma$ as fit parameters representing the flux at the decorrelation energy\footnote{The decorrelation energy corresponds to the energy at which the correlation between flux normalization and spectral index is minimum. The calculation of this energy is based on formula [3] in \cite{abdo10}.} E$_{\rm{dec}}$ and the  spectral index, respectively, for the observed ($\Gamma_{\rm{obs}}$) and intrinsic spectrum ($\Gamma_{\rm{int}}$). The fit parameters are listed in Table~\ref{tab:spectr_table_magic}. The observed spectra are quite soft, with a spectral index softer than 2, and in {the} case of RBS~0723 reaching the value $3.60 \pm 0.79$, where the error is statistical only.

For the sources without a detection (or hint-of-signal) in VHE gamma rays, flux upper limits were calculated (see Table\,\ref{tab:obs_table_magic}). Given their low redshifts and assuming that their VHE gamma-ray spectra were similar to that of the prototype EHBL 1ES\,0229+200, an observed photon index of 2 was  adopted for the upper limit calculations. For some of the sources, different photon indices (2, 3, and 4) were assumed to check the robustness of the upper limits.  In all cases, the calculated upper limits show small variations when different photon indices are assumed. However, these variations are within the instrument systematic uncertainties ($<15\,\%$). 
Given the VHE gamma-ray detection of 1ES\,1426+428 in 2012, the observed photon index of 2.6 was used for the calculation of the upper limits for the observation periods in 2010 and 2013, when the source was not detected.

\begin{table*}
\centering
\caption{\label{tab:spectr_table_magic} Results of the MAGIC spectral analysis of the EHBLs detected at VHE gamma rays {together with the one which shows hint of signal (RGB~J2042+244) and 1ES~0229+200, the reference source}.  {Columns from \textit{left} to \textit{right}}: Source name, redshift, decorrelation energy, differential energy flux derived from the observed spectrum at the decorrelation energy,  spectral index of the observed spectrum,  spectral index of the intrinsic spectrum corrected for EBL absorption with the \citet{2008AA...487..837F} model. Only statistical errors are reported.}
\begin{tabular}{lccccc} 
\hline
\multirow{2}{*}{Source}      & \multirow{2}{*}{$z$} & $E_{\mathrm{dec}}$ & $F_0$ $\times 10^{-12}$& \multirow{2}{*}{$\Gamma_{\mathrm{obs}}$} & \multirow{2}{*}{$\Gamma_{\mathrm{int}}$}\\
               & &[GeV]  &[cm$^{-2}$s$^{-1}$] & & \\
\hline

TXS~0210+515   & 0.049 & 1574 & $0.10 \pm 0.03$ & $2.0 \pm 0.3$ & $1.6 \pm 0.3$\\
RBS~0723       & 0.198 & 300 & $10.0 \pm 2.0$  & $3.6 \pm 0.8$ & $2.7 \pm 1.2$ \\
1ES~1426+428$^{\star}$ & 0.129 & 242 & $25.6 \pm 0.1$  & $2.6 \pm 0.3$ & $1.8 \pm 0.5$\\  
1ES~2037+521   & 0.053& 400 & $5.6 \pm 0.6$ & $2.3 \pm 0.2$ &  $2.0 \pm 0.5$\\
RGB~J2042+244$^{\dagger}$  & 0.104 & 379 & $2.6 \pm 0.5$ & $2.3 \pm 0.3$ & $ 1.7 \pm 0.6$ \\
\hline
1ES~0229+200   & 0.140 & 521  & $3.6 \pm 0.4$ & $2.6 \pm 0.1$ & $1.8 \pm 0.1$ \\
\hline
\end{tabular}
\begin{tablenotes}
\item{$\star$ Data from 2012 sub-sample.}
\item{$\dagger$ Only hint of signal was detected for this source.}
\end{tablenotes}
\end{table*}

In order to evaluate and compare the intrinsic emission of each source, the observed spectra have been corrected for the EBL absorption assuming the model by \citet{2008AA...487..837F}, filled black markers. The indices are reported in Table~\ref{tab:spectr_table_magic}, last column, where the errors listed are statistical only. 

\citet{2019MNRAS.486.4233A} tested the effect of using eight different EBL models, including those described by \citet{2008AA...487..837F} and \citet{2011MNRAS.410.2556D}, on the EBL density constraints. Their results show that such an effect is negligible within the tested models.

Very remarkably, the intrinsic spectral indices obtained by fitting with a power law function (dashed blue lines in Fig.~\ref{fig:MAGICSpectra}) are all quite hard suggesting that the VHE gamma-ray emission covers the energy range still below the second, high-energy SED peak. RBS~0723 represents the only exception, even if the faintness of the signal combined with the large distance severely affect the observed and de-absorbed spectra. 
Therefore, according to the MAGIC observations TXS~0210+515, whose intrinsic spectral index $\Gamma_{int}$ is $1.6\,\pm\,0.3$, is a newly detected  hard-TeV EHBL. 1ES~1426+428 and 1ES~2037+521, $\Gamma_{int}=1.8\,\pm\,0.5$ and $\Gamma_{int}=2.0\,\pm\,0.5$ respectively, are also compatible with the hard-TeV EHBL nature hypothesis. The hint-of-signal source RGB~J2042+244, $\Gamma_{int}=1.7\,\pm\,0.6$, seems also a hard-TeV EHBL. The extreme position of the second peak in these sources will be further investigated in Section~\ref{sec:modelling}. 

\section{\label{sec:HE}{\it Fermi}-LAT results}

{In general, EHBLs are not strong sources in the HE gamma-ray domain. The shift of the IC peak position to higher energies, together with the average low luminosity of these objects, make them faint sources for {\it Fermi}-LAT below 100\,GeV.}

For the determination of the HE gamma-ray properties of the sources of this study, the analysis of {\it Fermi}-LAT data was performed. The details of the analysis are reported in Appendix~\ref{app:fermi_analysis}.

The time span selected for each analysis varies in function of MAGIC exposure and source faintness. For each source the interval was selected as short as possible to match the MAGIC observations  to gather a TS $>$ 25. Taking into account the low fluxes involved, the minimum interval considered was as long as 1 year. 

In Table~\ref{tab:spectra_LAT}, last four columns, the main results of the analyses are reported. For comparison, the 3FGL, 2FHL \citep{Ackermann16}, and 3FHL \citep{ajello18} values are available in Appendix, Table~\ref{tab:src_lat}. 

Only one of the considered sources, namely RBS~0921, is not reported in any {\it Fermi}-LAT {catalog} yet. Interestingly, the analysis of more than 8 years of data from the source RBS~0921 indicates a TS of 23, corresponding to a significance of $\sim$4\,$\sigma$, near the threshold used to define a source detected at HE. The source therefore shows a hint of signal at HE with this deep exposure and will be possibly detected in the near future. All the other sources are detected with a TS spanning from 34, for the source RGB~J2313+147 (1 year exposure), to 94, for 1ES~1426+428 (1 year exposure) that is also the brightest source of the sample in X-ray. The fluxes measured in the 1 -- 300\,GeV energy range are between 1.4 to 6.7 $\cdot$ 10$^{-10}$ cm$^{-2}$s$^{-1}$. Therefore in this energy range the average integral flux of the sources lies within half order of magnitude. The  spectral index values are all below 2, which in the E$^2$dN/dE representation corresponds to an increasing spectrum. This is consistent with the extreme location of the second SED peak.

The {\it Fermi}-LAT spectral indices reported in  Table~4 are all compatible with the indices measured at higher energies with MAGIC, Table~3.  The similar indices are in agreement with the {behavior} observed in 1ES~0229+200, where the spectrum shows no break from the GeV up to the VHE range above 100\,GeV. However, in our case this compatibility could be simply due to the large error bars affecting the MAGIC determination (in particular for RBS~0723 and TXS~0210+515). Further, deep VHE measurements are needed to constrain the spectral shape of these EHBLs and determine with precision the location of the high-energy SED peak.

A study of the relation between the HE spectral properties and the TeV detectability, reported in Appendix~\ref{app:fermi_analysis}, reveals that there is no evident correlation between the measured LAT spectral index and the TeV detection.

\begin{table*}[ht!]
	\centering
		\caption{\label{tab:spectra_LAT}Main spectral parameters resulting from the analysis of {\it Swift}-XRT and {\it Fermi}-LAT data. Columns from \textit{left} to \textit{right}: source name; {\it Swift}-XRT observation dates (selected for the SED modeling); X-ray flux in the 2-10 keV energy range; spectral index of X-ray spectrum; fit-statistics parameters;  date for {\it Fermi}-LAT data (centered on the MAGIC observation window); HE gamma-ray flux in range of 1-300 GeV; spectral index of HE gamma-ray spectrum; likelihood test statistics (TS) of the fitted model.}
 		
 		\setlength{\tabcolsep}{0.36em}
 		\begin{tabular}{lcccc|cccc} 
 			\hline
 			       &\multicolumn{4}{c|}{{\it Swift}-XRT} & \multicolumn{4}{c}{{\it Fermi}-LAT}  \\ 
 		    \cline{2-9}
 			Source & Obs. date & F$_{(2-10\,\text{keV})}\times$10$^{-12}$  & \multirow{2}{*}{$\Gamma$} & \multirow{2}{*}{$\chi^2$/d.o.f.} & Interval & F$_{(1-300\,\text{GeV})}\times$10$^{-10}$ & \multirow{2}{*}{$\Gamma$} & \multirow{2}{*}{TS}$^\dagger$  \\
                    & [MJD] & [erg cm$^{-2}$s$^{-1}$]& & & [MJD]   & [cm$^{-2}$s$^{-1}$]   & & \\
 			\hline
 	        TXS~0210+515 & 57417 & $8.6 \pm 0.4$ & $1.71 \pm 0.04$&  119.4/77  & 57388--58118 & 4.3 $\pm$ 1.3   & 1.8 $\pm$ 0.2  &  42 \\ 
 	         TXS~0637-128 & 57784 & $15.6 \pm 1.0$ & $1.96 \pm 0.07 $ & 32.1/32  & 54682--58318 & 3.4 $\pm$ 1.1  & 1.5 $\pm$ 0.2 &  60\\ 
            BZB~J0809+3455 & 57126 & $2.1 \pm 0.3$ & $1.89 \pm 0.08$ & 9.5/17 & 56658--57753 & 2.4 $\pm$ 0.8    & 1.9 $\pm$ 0.2 &  39\\
            RBS~0723 & 57671 & $13.0 \pm 0.7$ & $1.68 \pm 0.04$ & 55.3/54 &  56108--57203& 2.8 $\pm$ 0.8   & 1.6 $\pm$ 0.2 &  53 \\
            1ES~0927+500  & 55648 & $6.4 \pm 0.7$ & $2.06 \pm 0.07$& 38.8/26 & 55562--57022 & 1.4 $\pm$ 0.6    & 1.5 $\pm$ 0.2 &  30   \\
            RBS~0921   & 57434 & $4.2 \pm 0.6$  & $1.63 \pm 0.09$ & 10.7/14   & & -      & - &  23 \\
            1ES~1426+428 & 56064 & $47.4 \pm 1.4$ & $1.84 \pm 0.02$ & 171.2/172  & 55927--56292& 6.7 $\pm$ 1.7   & 1.4 $\pm$ 0.2 & 94 \\
            1ES~2037+521$^\star$ & 57660 & $10.7 \pm 1.0$ & $1.93 \pm 0.13$ & 18.7/17 & 57203--57934& 4.6 $\pm$ 1.5  & 1.7 $\pm$ 0.2  & 46\\
            RGB~J2042+244 & 57192 & $9.2 \pm 0.8$ & $1.93 \pm 0.07$ & 29.5/27& 56838--57569 & 4.6 $\pm$ 1.4   & 1.7 $\pm$ 0.2 & 58  \\
            RGB~J2313+147 & 57172 & $1.6 \pm 0.1$ & $2.18 \pm 0.06$ & 30.5/32 & 56838--57569 & 3.6 $\pm$ 1.1  & 1.7 $\pm$ 0.2 & 34 \\
 			    \hline
 			1ES~0229+200  & 56264 & $13.1 \pm 1.0$ & $1.79 \pm 0.07$& 43.5/41 & 56293--58118 & 2.3 $\pm$ 0.7  & 1.5 $\pm$ 0.2 & 78\\
 			\hline
	 	\end{tabular}
	 	\begin{tablenotes}
         \item $\star$ The X-ray energy range for spectral analysis is 1.5--10 keV (see {Appendix}~\ref{app:xrt_analysis} for details).
         \item $\dagger$ {The square root of the TS is approximately equal to the detection significance for a given source.}
        \end{tablenotes}
\end{table*}

\section{\label{sec:X-ray_prop}X-ray properties of the sample}
EHBLs are, by definition, characterized by a synchrotron peak energy exceeding 10$^{17}$ Hz. This means that the bulk of the synchrotron emission is located in the X-ray band. For this reason special attention has been paid to the X-ray data for the study of the characteristic emission from the selected targets, in particular to those collected with the the X-ray Telescope (XRT) \citep{Burrows04} on-board of the {\it Neil Gehrels Swift Observatory}, and the Nuclear Spectroscopic Telescope Array ({\it NuSTAR}). 

\subsection{{\it Swift}-XRT results}
When possible, {\it Swift}-XRT data simultaneous with MAGIC pointings were requested via Target of Opportunity (ToO) observations. Moreover, all the available {\it Swift}-XRT archival data \citep{2013ApJS..207...28S} have been analyzed {using the procedure detailed in Appendix~\ref{app:xrt_analysis}.}

The X-ray light-curves of the targets in the 2 to 10 keV energy range are shown in the left panels of Figure~\ref{Fig:Xray_LC}. An example of the results is shown in Appendix~\ref{app:xrt_analysis}, Table~\ref{tab:obs_table_swift2}. For all the sources, the spectral index $\Gamma$ of the power law fitting the spectrum is {almost} $\lesssim$2. This indicates that the synchrotron peak lies around or above this energy range, as expected for this class of sources. The only exception is RGB~J2313+147, whose X-ray data 
suggest a peak located below 10$^{17}$\,Hz (see Sec.~\ref{subsec:nustar}).

For broad band SED modeling of each object, we selected the {\it Swift}-XRT observation which is either simultaneous to {\it NuSTAR} observations (TXS~0210+515, RGB~J2313+147, and 1ES 0229+200) or has the lowest time lag from the strongest detected signal in VHE gamma-ray band (Table~\ref{tab:spectra_LAT}). 

\begin{figure*}
\includegraphics{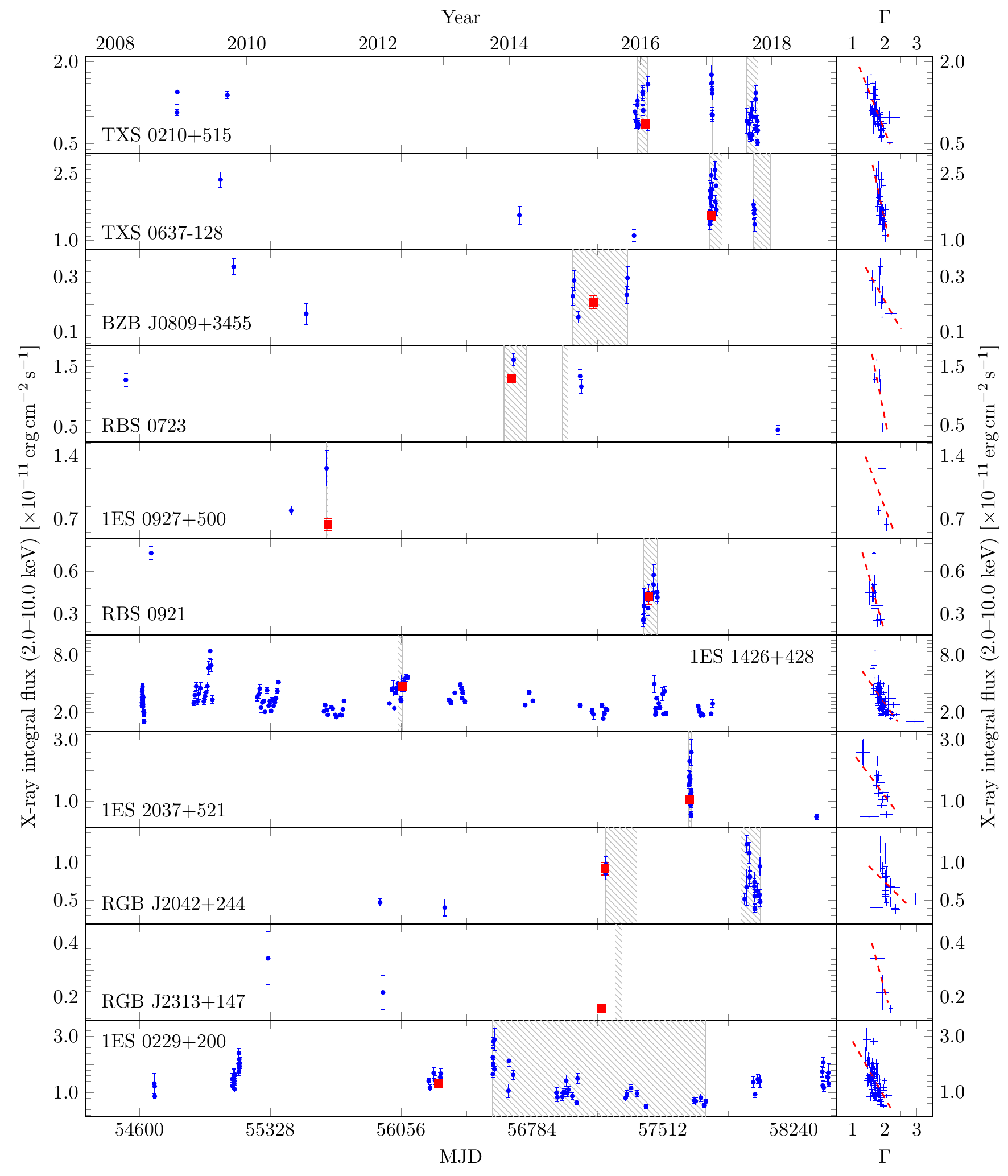}
\caption{\label{Fig:Xray_LC}\textit{Left panels:} X-ray light curve (2-10\,keV), corrected for Galactic extinction of the sample. The red {squares} show the data point which are used in broadband SED modeling. Shadowed areas illustrate MAGIC observation windows. \textit{Right panels:} Scatter plot of the power-law photon index ($\Gamma$) versus X-ray flux (2-10\,keV) measured with {\it Swift}-XRT for each source of the sample. Dashed lines are the best-fitted linear models fitted to the data of each source.}
\end{figure*}

As shown in the right panels of Figure~\ref{Fig:Xray_LC}, the possible relation between $\Gamma$ and the flux in the 2-10\,keV energy band is investigated for each source. The general trend is a harder-when-brighter {behavior}, meaning that the photon index decreases when the flux increases. This trend is quite typical in blazars, and has been observed in several X-ray campaigns of Mrk~501 \citep{1998ApJ...492L..17P}.  Mrk~501 is one of the best sampled BL Lac objects, and it showed an EHBL {behavior} during some observational campaigns \citep{2018AA...620A.181A}. The observed trend can be interpreted as the emerging of an additional population of accelerated electrons in the jet during high-activity states.

It is important to note, however, that there are also counter-examples to this trend, such as  the observation campaign on Mrk~501 in 2012, when the source exhibited very hard spectra in the X-ray and VHE ranges both in a quiescent and a flaring state \citep{2018AA...620A.181A}. This underlines the overall complexity of blazars when studied in detail.

\subsection{{\it NuSTAR} results}\label{subsec:nustar}

{\it NuSTAR} \citep{harrison13} observed TXS~0210+515 and RGB~J2313+147 in the hard X-ray band (3-79\,keV) with its two coaligned X-ray telescopes with corresponding focal planes, focal plane module A (FPMA) and B (FPMB), on 2016 January 30 and 2015 May 30, for a net exposure time of 21.4\,ks and 22.9\,ks, respectively.

{\it NuSTAR} data of TXS~0210+515 and RGB~J2313+147 have been processed as reported in Appendix~\ref{app:nustar_analysis}. Simultaneously to {\it NuSTAR} observations, {\it Swift}-XRT observations of TXS~0210+515 and RGB~J2313+147 were performed. This allows us to study the X-ray spectra of each source over a wide energy range. The results of the simultaneous fits of the {\it NuSTAR} and {\it Swift}-XRT data are presented in Appendix, Table~\ref{Tab:nustar0210and2313}. All errors are given at the 90\% confidence level. The photoelectric absorption model \texttt{tbabs}, with a neutral hydrogen column density fixed to its Galactic value was included in all fits. To account for the cross-calibration between {\it NuSTAR}-FPMA, {\it NuSTAR}-FPMB, and {\it Swift}-XRT a constant factor was included in the model, frozen at 1 for the FPMA spectra and free to vary for the FPMB and XRT spectra. The difference of the cross-calibration for the FPMB spectra with respect to FPMA spectra is 1-3 percent, while for the XRT spectra is $\leq 10\%$ and $\leq 15\%$ in the case of TXS~0210+515 and RGB~2313+147. \citet{2017AJ....153....2M} claimed that the relative quality of the spectra play significant role in calculation of cross-normalization constant between the two instruments. The difference of the cross-calibration for the XRT spectra with respect to FPMA is in agreement with their finding.

Two different models were tested: a simple power law and a log parabola model. For TXS~0210+515, the F-test shows an improvement of the fit with a log parabola model with respect to a simple power law, with a probability that the null hypothesis is true of 9.8$\times$10$^{-9}$. The log-parabola model is therefore preferred with $5.7 \sigma$ level of confidence. The combined {\it Swift}-XRT and {\it NuSTAR} spectrum of TXS~0210+515 is reported in the Appendix, Figure~\ref{Fig:0210and2313_XRT_NuSTAR}. 

In the case of RGB~2313+147, the X-ray spectrum is well fitted by a simple power law (Fig.~\ref{Fig:0210and2313_XRT_NuSTAR}). However, the X-ray flux observed during the {\it NuSTAR} observation of RGB~2313+147 is a factor of 10 lower with respect to the value observed for TXS~0210+515. In this way, the relatively low number of counts may prevent us from accurately test a curved spectrum in X-rays.

1ES~0229+200 was also observed with {\it NuSTAR} on 2013 October 02, 06, and 10, for a total exposure time of $\sim$ 51\,ks. We adopt here the data analysis results published in \citet{2018MNRAS.477.4257C}. Also in this case a  log parabola model is statistically preferred  over a simple power law model.

\section{\label{sec:OtherBands}Properties of the sample in other bands}

All the ten targets considered in the study have radio data accessible via public archives that were recovered from the NED database\footnote{\href{https://ned.ipac.caltech.edu}{https://ned.ipac.caltech.edu}}. The apparent radio flux values measured at 1.4 GHz distribute from 4 to 500\,mJy. The corresponding absolute powers distribute in the range $(1 - 6) \times 10^{33}$\,W.

The XRT data presented in previous Section have always been complemented with data at lower frequencies collected with the UVOT instrument, onboard the {\it Swift} satellite. Apart from the bands at larger energies, in the UV domain (when available), the UVOT data generally represent the emission from the host galaxy. In extreme blazars, in fact, the host galaxy is clearly detected at IR-optical wavelengths, as the synchrotron peak is shifted towards the X-ray regime. This is not the usual case for other kind of BL Lac objects, where the host galaxy is usually dominated by the peak of the non-thermal continuum.

Five sources of the sample are reported in the {\it Swift}-BAT 105-Month Hard X-ray catalog\footnote{\href{https://swift.gsfc.nasa.gov/results/bs105mon/}{https://swift.gsfc.nasa.gov/results/bs105mon/}}, they are TXS~2010+515, TXS~0637-128, 1ES~0927+500, 1ES~1426+428, and 1ES~0229+200 \citep{swiftbat}. Interestingly, three of those sources have been detected by MAGIC {suggesting that the detection in hard X-rays is a good (but not exclusive) selection criterion for VHE observations.}

\section{\label{sec:modelling}SED modeling}

The SEDs of each target are assembled complementing the MAGIC, {\it Swift}-XRT,  {\it NuSTAR}, and {\it Fermi}-LAT data with archival data from the ASI Space Science Data Center (SSDC)\footnote{\href{http://www.asdc.asi.it}{http://www.asdc.asi.it}}.
VHE gamma-ray data are corrected for the EBL absorption effect by adopting the \citet{2008AA...487..837F} model, which is in good agreement with current limits for the diffuse background  \citep{2016RSOS....350555C}.  

The SEDs are displayed in Figure~\ref{fig:SEDs_detection}. The archival data are shown in {gray} while the data used for the modeling are displayed with red open markers and red downward triangles in case of upper limits. These data can be considered as quasi-simultaneous, with MAGIC and {\it Fermi}-LAT data being integrated over a long period due to the relatively faint emission, and {\it Swift}-XRT and {\it NuSTAR} spectra taken from one observation within the MAGIC observation window. For 1ES~0229+200 the {\it NuSTAR} data recently published in \citet{2018MNRAS.477.4257C} were adopted. In the case of 1ES~1426+428 the average 14-195 keV spectrum obtained with {\it Swift}-BAT in 105 months of survey from 2004 to 2013 \citep{swiftbat} was included in the archival (gray) SED and clearly constrain the peak position in the extreme region, above 10$^{17}$\,Hz. 

\subsection{SSC model}
For fitting the broadband spectra, first the numerical code in \citet{2014ApJ...780...64A} {\citep[see also][]{2015ApJ...808L..18A,asano18}}, which calculates the emission from a conical jet, is  adopted. In this code, the temporal evolution of the electron and photon energy distributions in the plasma rest frame are calculated along the jet. In the steady outflow scenario, the temporal evolution along the jet is equivalent to the radial evolution, so that the emission in this code is obtained from the integral of the 1-D structure. This treatment is similar to the {\tt BLAZAR} code by \citet{2003AA...406..855M}, which has been frequently adopted to reproduce blazar spectra \citep[see e.g.,][]{2008ApJ...672..787K,2012ApJ...754..114H}. The conically expanding jet naturally leads to adiabatic cooling of electrons, which is a similar effect to the electron escape in one-zone steady models. Thus, the electron escape in this 1-D code can be neglected. 

The injection of the non-thermal electrons starts from an initial radius $R$ = $R_0$. The electron injection is assumed to continue during the dynamical timescale $R_0/(c \Gamma)$ in the plasma rest frame. In this timescale, the injection rate into a given volume $V$, which is expanding as $V \propto R^2$, is assumed to be constant. Even after the shutdown of the electron injection, the electron energy distribution and photon emission is calculated as far as $R$ = 10 $R_0$. The injection rate is normalized by the electron luminosity $L_{\rm e}$ in the observer frame. The electron energy distribution at injection is a single power law with an exponential cutoff, $\dot{N}(\gamma) \propto \gamma^{-p_1} \exp(-\gamma/\gamma_{\rm max})$ for the electron Lorentz factor $\gamma>\gamma_{\rm min}$, or a broken power-law energy distribution, changing the index from $p_1$ to $p_2$ at $\gamma = \gamma_{\rm br}$. The magnetic field in the plasma frame evolves as $B$ = $B_0 (R_0/R)$ in the code. Synchrotron, IC scattering with the Klein--Nishina effect, $\gamma \gamma$-absorption, secondary pair injection, synchrotron self-absorption, and adiabatic cooling are taken into account.

In this paper, the jet opening angle is assumed to be 1 /  $\Gamma$, where $\Gamma$ is the bulk Lorentz factor of the jet, and  an on-axis observer (the viewing angle is zero) is considered. The photon flux is obtained by integrating over the entire jet, taking into account the Doppler boosting by the conically outflowing emission region.

The data cannot constrain all the model parameters. Here, the initial radius is fixed at a typical value being $R_0$ = 0.03 pc, and the minimum Lorentz factor at \mbox{$\gamma_{\rm min}$ = 20}. The remaining 5 model parameters, i.e, $\Gamma$, $B_0$, electron luminosity $L_{\rm e}$, maximum electron Lorentz factor $\gamma_{\rm max}$, and spectral index $p_1$  are left free to vary. The broken power-law model includes two additional parameters, that is the break Lorentz factor $\gamma_{\rm br}$ and the high-energy spectral index $p_2$. The parameters in the fits are {summarized} in Table~\ref{param_tab} together with the values obtained from the fits: the synchrotron peak {frequency} ($\nu_{\rm syn,pk}$), the IC peak {frequency} ($\nu_{\rm IC,pk}$), the Compton dominance parameter (the ratio of $\nu L_\nu$ at $\nu_{\rm syn,pk}$ to that at $\nu_{\rm IC,pk}$, dented as ``CD''), and the energy density ratio of the magnetic field with that of the electrons ($U_B/U_{\rm e}$) at the radius where the electron injection terminates. 

Note that the Klein--Nishina effect is crucial in EHBLs. If we can use the well-known relation $\nu_{\rm IC,pk} \sim \gamma_{\rm max}^2 \nu_{\rm syn,pk}$ or $\nu_{\rm IC,pk} \sim \gamma_{\rm br}^2 \nu_{\rm syn,pk}$ in the Thomson regime, the parameter estimate is straightforward. However, the photon energy in the electron rest frame is much higher than $m_{\rm e} c^2$ in EHBLs, so that the simple estimate for $\nu_{\rm IC,pk}$ is not useful because of the Klein--Nishina effect. Our numerical code, which includes the Klein--Nishina effect, outputs a consistent magnetization, which is much less than the Compton dominance parameter introduced above.

\begin{figure*}
\begin{center}
\includegraphics[width=1.0\textwidth]{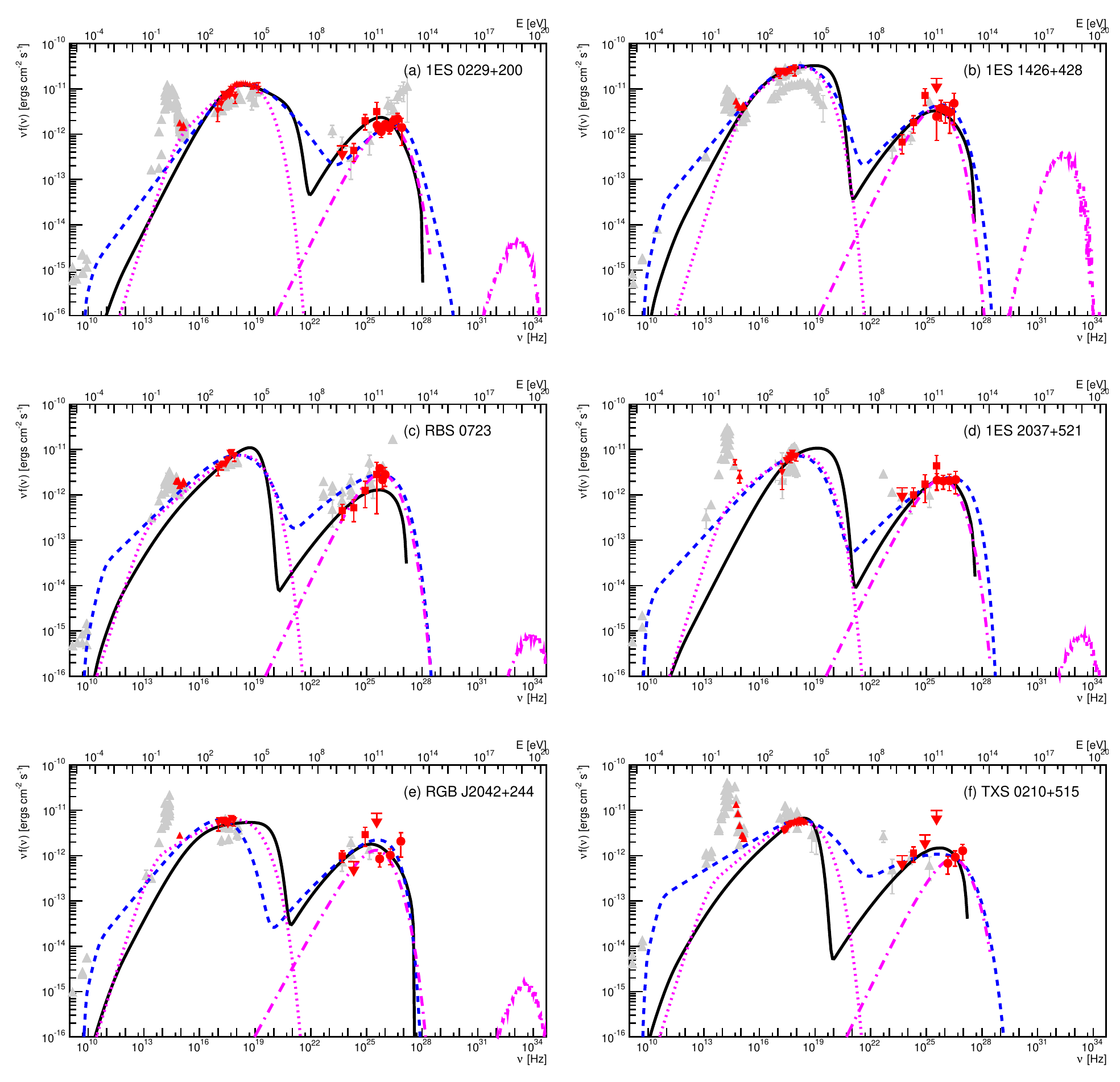}
\end{center}
\vspace{-0.7cm}\caption{\label{fig:SEDs_detection}Broadband SED and modeled spectrum for 1ES~ 0229+200 (archetipal EHBL) and the {four VHE gamma-ray sources detected with MAGIC presented in the study. The broadband SED of RGB J2042+244, for which a hint of signal was detected at VHE gamma-rays  is also shown}. Red points represent contemporaneous UVOT, XRT, {\it NuSTAR}, {\it Fermi}-LAT and MAGIC data considered in the fit. Gray markers are archival data from ASDC website. Blue, dashed line is the result of the  conical-jet SSC model. Black continuous line represents the outcome of the spine-layer model. Dashed-dotted magenta line is the outcome of the proton-synchrotron model. The third bump in the proton-synchrotron model is the expected neutrino flux resulting from the best fit solution proposed. Details in the text.}
\end{figure*}

First, we consider 1ES~0229+200, the prototype of EHBLs.  As shown in Figure~\ref{fig:SEDs_detection}~(a), the {\it NuSTAR} data provide the spectral shape around the synchrotron peak very well. This sharp break cannot be reproduced by the cooling break, so that the broken power-law injection is adopted. The model {is in a good agreement with the  observed  quasi-simultaneous data.} Assuming the synchrotron radiation is the dominant cooling process, the cooling break in the electron energy distribution is expected to appear at

\begin{eqnarray}
\gamma_{\rm c}=
\frac{6 \pi m_{\rm e} c^2 \Gamma}{\sigma_{\rm T} B^2 R_0},
\end{eqnarray}
This corresponds to an observed photon energy
\begin{eqnarray}
\varepsilon_{\rm syn,c}&=&\frac{3}{2} \Gamma \frac{\hbar e B}{m_{\rm e} c}
\gamma_{\rm c}^2 \\
&\simeq& 8.7 \left( \frac{\Gamma}{20} \right)^3
\left( \frac{B}{0.1~\mbox{G}} \right)^{-3}
\left( \frac{R_0}{0.03~\mbox{pc}} \right)^{-2}~\mbox{keV}.
\end{eqnarray}
In the modeled spectrum, the break energy at $\sim 10$ keV due to $\gamma_{\rm br}$
and the cooling break at $\sim 300$ keV are consistent with a
magnetic field of 0.03 G at the radius where the electron injection terminates.
The {magnetization} parameter $U_{\rm B}/U_{\rm e}$
is very low ($\sim 10^{-3}$) in this model.

The MAGIC data show a significantly dimmer and softer spectrum than those observed in 2005--2006 by H.E.S.S. \citep{aharonian07}. Taking into account the H.E.S.S. data for the one-zone synchrotron self-Compton (SSC) model by \citet{2011AA...534A.130K} requires a very narrow electron energy distribution ($\gamma_{\rm min} = 3.9 \times 10^5$, $\gamma_{\rm br}$ = 6.2$\times 10^7$). While the size of the emission region in their model is only by a factor of $\sim 2$ larger than ours, their magnetic field is much lower ($3.2\times 10^{-5}$ G). \citet{2018MNRAS.477.4257C} also fitted the broadband spectrum of this source, adopting the same X-ray data as included in our modeling, but using the H.E.S.S. data. With $U_B/U_{\rm e}=5.9 \times 10^{-6}$ -- $5.0 \times 10^{-5}$ the {magnetization} parameter in their models is also extremely low. However, with the shown mild variability in VHE gamma rays of 1ES 0229+200 \citep[VERITAS, ][]{aliu14} and the non simultaneity of the {\it Swift} and H.E.S.S. data the modeling can be affected. The fitting result with the MAGIC data agrees with a more conservative electron energy distribution and {magnetization}.

The synchrotron spectral peak for 1ES~1426+428 is not well constrained by the data collected during the MAGIC observing period (see Fig.~\ref{fig:SEDs_detection}(b)). Referring to the historical data, a single power-law injection model with the peak energy $\varepsilon_{\rm syn,pk}\sim 6$ keV is adopted in that figure. In this case, a larger magnetic field is adopted, implying that the synchrotron peak is due to the cooling break. The broad shape of the synchrotron peak leads to a relatively higher photon flux in the lower energy range. When the Klein-Nishina effect  becomes crucial, the higher density of low energy photons enhances the efficiency of SSC emission. The relatively broad spectral peak and different IC peak energies in 1ES~1426+428 lead to a large difference in the magnetization parameter even for a Compton dominance parameter similar to that of 1ES~0229+200.

Compared to the synchrotron spectral shape, the observed gamma-ray spectrum is very hard. Thus, the model has difficulty in reproducing the hard Fermi spectrum. Here, we give weight on the MAGIC data points, and the broadband spectrum is fitted.

For RBS~0723 (Fig.~\ref{fig:SEDs_detection}, c)  {compared to the synchrotron spectral shape, the observed HE gamma-ray spectrum is very hard. Thus, the model has difficulty in reproducing the hard {\it Fermi}-LAT spectrum. Here, we give weight on the MAGIC data points, and the broadband spectrum is fitted.} The single power-law injection model reproduces the synchrotron and SSC flux in the VHE band, while the {\it Fermi}-LAT flux lies below the model expectations. The synchrotron spectral peak is adjusted by the maximum electron energy. The cooling break is higher than $\varepsilon_{\rm syn,pk}$ in this case. The IC flux of the modeled spectrum is slightly higher than the {\it Fermi} flux, but consistent with the flux in other observational periods (in gray). 

The hard X-ray spectrum in 1ES~2037+521 indicates a peak energy higher than 4\,keV. The model shown in Figure~\ref{fig:SEDs_detection}(d) assumes the synchrotron peak to be determined by the electron maximum energy. Since the synchrotron peak is not constrained, we can increase $\varepsilon_{\rm syn,pk}$ with a larger $\gamma_{\rm max}$, which leads to further low magnetic field. The obtained magnetization in 1ES 2037+521 is the lowest among our results. Adopting a higher magnetic field, the break appears below 4\,keV. Among the models presented in this paper, 1ES~2037+521 has the highest $\varepsilon_{\rm syn,pk}$, which is close to 100\,keV.  This is much higher than the highest value ($\sim 9$\,keV) confirmed for BL Lacs in the steady state \citep{2018MNRAS.477.4257C}. The flat spectrum obtained with MAGIC seems consistent with the SSC peak of the modeled spectrum.

Assuming that the flat X-ray spectrum in RGB~J2042+244 corresponds to the synchrotron peak, the spectrum is fitted adopting a relatively lower value for the maximum energy of electrons as shown in Figure~\ref{fig:SEDs_detection}(e).

The synchrotron peak in TXS 0210+515 is relatively well constrained. To reconcile the flat gamma-ray spectrum, especially for the Fermi data, we need to assume a soft electron energy distribution as $p_1=2.5$, which implies that the energy budget is dominated by low energy electrons. As a result, the magnetization is one of lowest as $\sim 10^{-4}$.

\begin{figure*}
\centering
\includegraphics[width=1.0\textwidth]{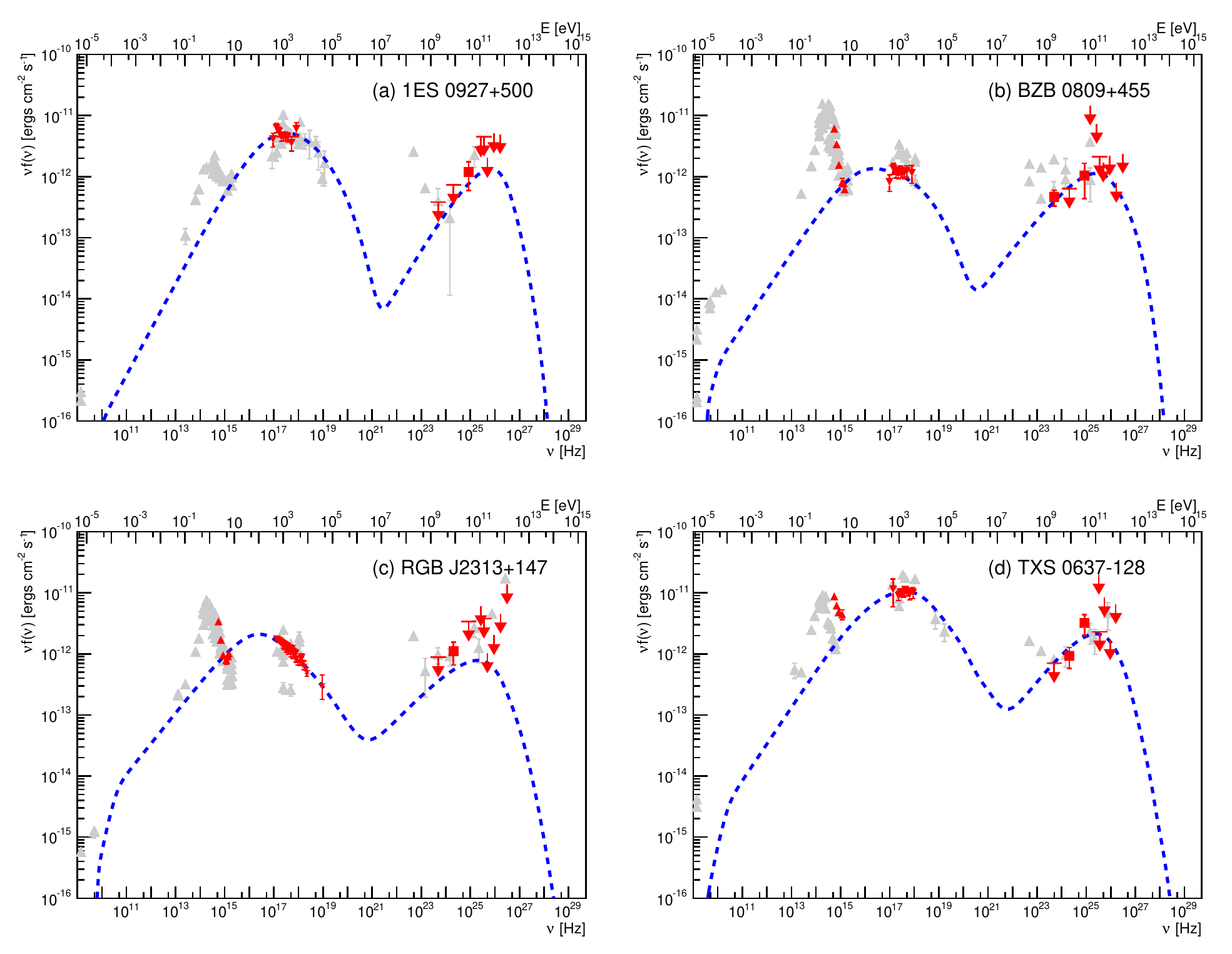}
\caption{\label{fig:SEDs_undetection}Broadband SED and modeled spectra for four sources with no MAGIC detection or hints (but {\it Fermi}-LAT detection) presented in the study. Red points represent contemporaneous UVOT, XRT, NuSTAR, {\it Fermi}-LAT and MAGIC data considered in the fit. Gray markers are archival data from SSDC website. Blue, dashed line is the result of the conical-jet SSC model. Details in the text.}
\end{figure*}

There are five sources for which MAGIC provides only upper limits in VHE flux. Even in these cases, the upper limits can constrain the model. In 1ES~0927+500, there are significant  upper limits at roughly 600\,MeV and 200\,GeV by {\it Fermi} and MAGIC, respectively, while the source was detected around 100\,GeV (Fig.~\ref{fig:SEDs_undetection}(a)).
To fit the spectrum without taking into account the MAGIC upper limits, the bulk Lorentz factor is adjusted to 10, while a hard electron spectrum ($p_1$ = 1.5) needs to be assumed to avoid the {\it Fermi} upper limits.

The MAGIC upper limits between 200 and 700\,GeV constrain well the modeled spectrum for BZB~J0809+3455 (Fig.~\ref{fig:SEDs_undetection}(b)). In this case, the model suggests that the synchrotron peak energy is below the peak energy criterion for EHBLs. 

The soft X-ray spectrum in RGB~J2313+147 (Fig.~\ref{fig:SEDs_undetection}(c)) also implies that this falls not into the EHBL classification. The fitting result constrained by the MAGIC upper limits leads to $\varepsilon_{\rm syn,pk}\simeq 100$ eV. 

For TXS~0637-128, we adopted the redshift $z=0.136$ for our modeling (S. Paiano, private communication). The synchrotron spectral peak is produced by the electron cooling effect. The magnetization is the highest in our model samples, Figure~\ref{fig:SEDs_undetection}(d).

The upper limits in the VHE range for RBS~0921  do not sufficiently constrain the model, therefore the modeling of the broadband spectrum is omitted in this case. The SED is reported in Appendix, Figure~\ref{fig:0921_SED}.

To summarize, the hard gamma-ray spectra seen in 1ES~0229+200, 1ES~1426+428, and 1ES~2037+521 were reproduced consistently with the spectral shape of the synchrotron component.  Three different mechanisms were considered in the samples to form the synchrotron peak: the intrinsic break in the electron spectrum (1ES~0229+200, 1ES~0927+500, BZB~J0809+3455, RGB~J2313+147), the maximum electron energy (RBS~0723 and RGB~J2042+244), and the cooling break (1ES~1426+428 and 1ES~2037+521). In general, we find that EHBLs have high values for $\gamma_{\rm br}$ or $\gamma_{\rm max}$ and a high synchrotron {peak frequency} $\nu_{\rm syn,pk}$, which implies the Klein--Nishina effect to be crucial. High-energy electrons interact mainly with photons with much lower {frequency} than $\nu_{\rm syn,pk}$. The flux ratio of the two spectral components in EHBLs seems not directly related to the {magnetization} parameter. According to the model, 1ES~0229+200 remains the source of the sample with the most extreme synchrotron peak, while RGB~J2313+147 and BZB~J0809+3455 are non-EHBL sources, having their peak below the defined threshold of 10$^{17}$\,Hz. Interestingly, the SED models of the remaining sources feature a synchrotron peak frequency in good agreement with the estimates of the 2WHSP reported in Table~\ref{tab:src_table} with the exception of RGB~J2313+147, whose peak was estimated at higher frequencies $\nu_{peak; 2WHSP}$ = 10$^{17.7}$\,Hz , $\nu_{peak; ssc}$ = 10$^{16.5}$\,Hz) and TXS~0210+515 whose SSC model predicts a much higher peak frequencies instead ($\nu_{peak; 2WHSP}$ = 10$^{17.3}$\,Hz , $\nu_{peak; ssc}$ = 10$^{18.3}$\,Hz).

In our sample, in spite of the divergency in the model, the {magnetization} parameters $U_B/U_{\rm e}$ are commonly small. A comparison can be performed with Mrk~421, one of the most precisely observed blazars, where the {magnetization} has been estimated as a few percent \citep{2011ApJ...736..131A,asano18}. The typical value of $\sim 10^{-3}$ found in the sample is much lower than that found in Mrk~421, implying a low magnetic field that is unfavorable for magnetic reconnection models \citep[see e.g.][and references therein]{2015MNRAS.450..183S}. This also raises contradiction with the magnetically driven jet model. Radio observations for the radio galaxy M~87 revealed that the radio core region is dominated by the magnetic energy \citep{2015ApJ...803...30K} and the bulk Lorentz factor and jet width profiles along the jet \citep{2013ApJ...775..118N} are consistent with a magnetically-driven parabolic jet model \citep{2009MNRAS.394.1182K}. These observations support highly {magnetized} jet models, but the spectra in EHBLs may require either a fast dissipation of the magnetic field at the root of the jet or another jet acceleration model.

It should be noted that large error bars permit to adopt different parameter sets. Therefore, $R_0$ was fixed to search for conservative parameters in this paper. The parameters in Table~\ref{param_tab} are such examples. Moreover, considering the short variability in blazars, the GeV--TeV fluxes obtained with long integration times are not completely simultaneous with observations at other wavelengths. These uncertainties may change the interpretation, especially for the {magnetization}. In fact in 1ES~2037+521, for example, another parameter set was found when implying $U_B/U_{\rm e} \sim 10^{-5}$ different from the model presented in Figure~\ref{fig:SEDs_detection}(d). However, an extreme parameter set  such as a very low magnetic field ($U_B/U_{\rm e} \ll 10^{-3}$) or a very high $\gamma_{\rm min}$ is not necessarily required to fit the EHBL spectra in this paper.

\begin{figure}
\centering
\includegraphics[width=0.47\textwidth]{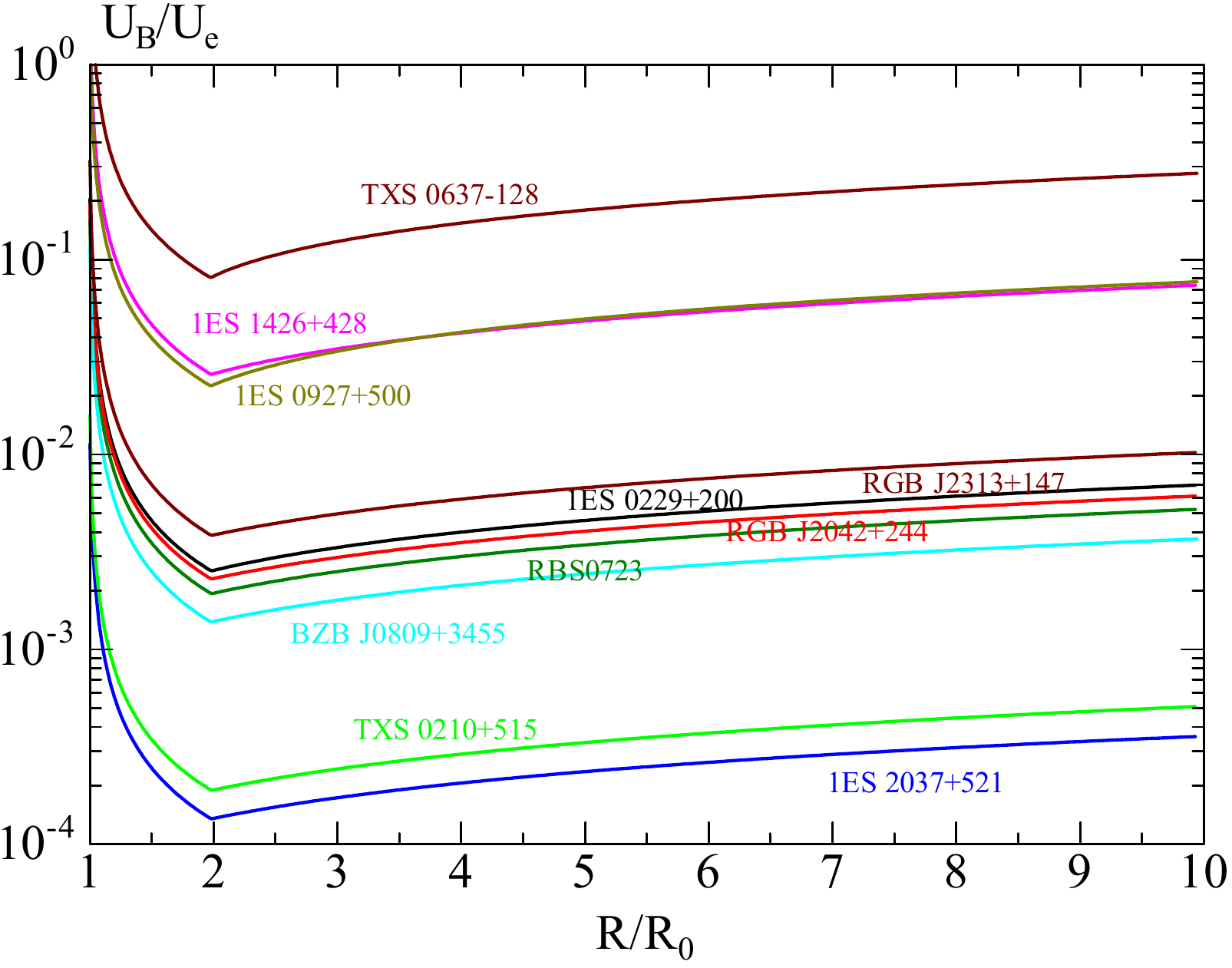}
\caption{\label{Fig:UBUE}U$_B$/U$e$ distribution as a function of R/R$_0$ for the sources considered in this study in the SSC model (see Tab.~\ref{param_tab}).}
\end{figure}
 		
\subsection{Spine Layer Model}
The main outcome of the modeling of the sample of EHBLs with the SSC model presented in the previous section is a rather low magnetization. This is somehow in contradiction with the theoretical and observational constrain of equipartition needed to launch and sustain the jet close to the central massive black hole. As discussed in \citet{tavecchio16}, a possibility to solve this problem is to decouple the synchrotron and IC components, assuming the existence of a supplementary source of soft photons intervening in the IC emission, as envisioned in the so-called spine-layer model \citep{2005AA...432..401G, tavecchio08}. In this model one assumes the existence of two regions in the jet: a faster inner core (the spine, with Lorentz factor $\Gamma$), surrounded by a slower sheath of material (the layer, with Lorentz factor $\Gamma_L$). The radiation emitted by one region as observed in the frame of the other is amplified because of the relative motion. In this way the IC luminosity of both components (in particular that of the spine) is increased with respect to that of the one--zone model. Given the larger radiation energy density with respect to the standard model, it is possible to increase the magnetic energy density (and decrease the electron energy density), thus reaching conditions close to equipartition. 

{In this scenario, the emission regions are filled with particles distributed in energy according to a smoothed broken power law:}

\begin{equation}
 N(\gamma)=K\gamma^{-n_1}\left(1+\frac{\gamma}{\gamma_{\mathrm b}}\right)^{n_1-n_2},   \gamma_{\mathrm min}<\gamma<\gamma_{\mathrm max},
\end{equation}

{The distribution has normalization $K$ between $\gamma_{min}$ and $\gamma_{max}$ and slopes $n_1$ and $n_2$ below and above the break, $\gamma_{\mathrm b}$ \citep{2003ApJ...593..667M}. T}his model requires to specify  a relatively large number of parameters. To reduce the free parameters, the Lorentz factors of the spine and the layer are fixed to $\Gamma=20$ and $\Gamma_L=3$ and the further assumption $\delta=\Gamma$ is made, thus fixing the viewing angle of the jet $\theta_v\simeq 2.9$ deg. {Moreover, the minimum electron Lorentz factor of spine is fixed to  $\gamma_{min}=100$}. The other parameters (in particular the luminosity of the layer emission) were varied so that the spine is close to equipartition.

This alternative scenario is tested on 1ES~0229+200 {as well as on the four sources with significant detection with MAGIC and RGB ~J2042+244, for which a hint of signal was found}. For the remaining sources, we notice that without a detection at VHE the parameters are not sufficiently constrained and therefore we do not further investigate the applicability of spine-layer (and {proton synchrotron}, see later) model. The results of the model are displayed in Figure~\ref{fig:SEDs_detection}, black continuous line. In Table~\ref{tab:spine}, the parameters used for the spine are reported. As expected, the values of the magnetic field adopted in this model are higher than those assumed in the SSC model and in all it is possible to obtain a satisfactorily fit of the data assuming rough equipartition conditions. Since equipartition also marks the condition to have the lowest jet power required to have a given radiative output (e.g. {\cite{GhisCel01}}), the jet powers estimated with the spine-jet scenario are systematically lower (by more than one order of magnitude) than those required by the SSC model.

\subsection{Proton synchrotron scenario} 
The second alternative model considered is a scenario in which proton synchrotron radiation is responsible for the $\gamma$-ray component of the blazar SED. Blazar hadronic emission models have long been considered a valid alternative to leptonic models, in particular thanks to the natural link they provide with neutrino astronomy and ultra-high-energy cosmic-ray acceleration in AGN jets. One weakness of blazar hadronic models is that they require a rather large power in the protons responsible for the emission, often larger than the Eddington luminosity of the black hole powering the AGN. This is particularly true for bright FSRQs, as discussed e.g. in \citet{zdziarski15}. For low luminosity BL Lacs, on the other hand, a proton synchrotron solution with a much lower, sub-Eddington, proton luminosity can be achieved, as discussed in \citet{cerruti15}. In addition, the absence of fast variability in EHBLs, in contrast with what observed in typical HBLs, is also consistent with the slow cooling time-scale of hadrons in the jet.

Similar to the spine-layer model case, the proton synchrotron model was tested only to the sources with a VHE gamma-ray spectrum determination. Without a spectral determination at VHE gamma rays, in fact, the proton-synchrotron component remains poorly constrained. Moreover, the number of free parameters of blazar hadronic models is much higher than the one of leptonic models, due to the extra proton energy distribution. In order to reduce the parameter space to study, some physically motivated assumptions are made:

\begin{itemize}
\item the Doppler factor of the emitting region $\delta$ is fixed to 30, a value typical for blazars  \citep{2010MNRAS.401.1570T}, and consistent with the estimates from radio observations.
\item the size of the emitting region $R$ is usually constrained by the observed variability time-scale via the usual causality argument; given that for the majority of the sources no fast (day-scale or less) variability is seen at any wavelength, a $R\leq 1.6\times10^{17}\ (1+z)^{-1} {\rm cm}$ is assumed. This value translates, for a Doppler factor $\delta=30$, into a variability time-scale of two days.
\item {minimum and break electron Lorentz factor is fixed to $\gamma_{e,\text{min}} = \gamma_{e,\text{break}}=200$. Minimum proton Lorentz factor is fixed to $\gamma_{p,\text{max}} =1$, while the break proton Lorentz factor ($\gamma_{p, break}$) is assumed to be equal to the maximum proton Lorentz factor ($\gamma_{p, max}$).}
\item the maximum proton Lorentz factor $\gamma_{p, Max}$ is constrained by equating acceleration and cooling timescales: the acceleration time-scale is expressed as $\tau_{acc} = \frac{m_p c}{\eta e B} \gamma_p$, where $\eta$ is a parameter defining the efficiency of the acceleration mechanism, fixed to $0.1$; the cooling time-scales considered are the adiabatic one, $\tau_{ad} \simeq \frac{2R}{c}$, and the synchrotron one.   
\item hadrons and leptons share the same acceleration mechanism, and in particular the power-law index of the injected particle distribution is identical, {i.e. $\alpha_{1} = \alpha_{e,1} = \alpha_{p,1}$ and $\alpha_{2} = \alpha_{e,2} = \alpha_{p,2}$}.
\item the lepton energy distribution at equilibrium is computed assuming that the main cooling mechanism is synchrotron radiation.
\end{itemize}

The proton synchrotron spectrum, with $\gamma_{p, Max}$ constrained as defined above, is {characterized} by a clear degeneracy in the $B$-$R$ plane, with solutions lying on a line $B \propto R^{-2/3}$ displaying the same peak frequency, being thus indistinguishable in absence of additional information (i.e. neutrinos, or on the basis of their proton power). It exists in addition a maximum peak frequency of the proton synchrotron component, which corresponds to the transition between adiabatic-dominated and synchrotron-dominated cooling regimes \citep[see][]{cerruti15}, and is equal to $1.28\times10^{26} \frac{1}{(1+z)} \frac{(3-\alpha_{p,1})}{1.5} \frac{\delta}{10} \textrm{Hz}$.

175 hadronic models are produced, scanning the following parameter space: $\nu_{syn,peak} \in [0.1\nu_{Max}, \nu_{Max}]$, $R\in [10^{14} \textrm{cm}, R_{Max}]$, and the proton normalization $K_p \in [K^\star/3, 3K^\star]$, where $K^\star$ corresponds to the proton density which provides a synchrotron spectrum at the level of the MAGIC spectra. Solutions which correctly describe the SED are selected via a $\chi^2$ test, identifying a posteriori the solution with the lowest $\chi^2$ and applying a $\Delta \chi^2$ cut corresponding to a 1-$\sigma$ interval. It is important to underline here that the $\chi^2$ is computed without taking into account systematic uncertainties on the spectral measurements of the various instruments. The corresponding model parameters are provided in Table~\ref{tab:hadronic-model}, while the minimum-$\chi^2$ proton-synchrotron solutions are shown in Figure~\ref{fig:SEDs_detection} together with the leptonic cases.

Proton synchrotron solutions provide a good description of the SEDs of extreme blazars, with luminosities which can be as low as $10^{45}$erg s$^{-1}$, only a small fraction of the Eddington luminosity of the super-massive black-hole powering the blazar, which is $1.26\times10^{47} (M/10^9 M_\odot)$ erg s$^{-1}$. One parameter which takes unusual values is the injection index of the particle distributions, which is very hard ($\alpha_{e,1} = \alpha_{p,1} = 1.1-1.3$) compared to the value expected from relativistic shock acceleration ($\alpha \simeq 2.2$). On the other hand, such hard values for the injection index can be compatible with particle acceleration by magnetic reconnection \citep[see e.g.][]{sironi14}. It is important to underline, however, that the values of $\alpha_{e,1} = \alpha_{p,1}$ are not the result of the SED modeling, but are a direct consequence of the hypotheses of co-acceleration of electrons and protons and of simple synchrotron cooling as the main driver for the steady-state electron distribution. Relaxing these hypotheses can lead to softer values for $\alpha_{e,1}$ and $\alpha_{p,1}$, more in line with shock acceleration.

In Figure~\ref{fig:SEDs_detection}, together with the electromagnetic emission, we also show the neutrino emission, which appears in the PeV-EeV band. The neutrino emission from all proton-synchrotron models is rather moderate, showing a typical peak flux several orders of magnitudes lower than the gamma-ray peak. 
While the proton-synchrotron model is degenerate in terms of photon emission, it predicts different neutrino fluxes as a function of the compactness of the emitting region (smaller and denser emitting regions resulting in a higher rate of proton-photon interactions, and thus neutrino production). The maximum neutrino flux expected from the proton-synchrotron models for the six sources under study is shown in Appendix~\ref{app:neutrinoSED}. The most promising source in terms of neutrino output is 1ES~1426+428, which due to the bright soft photon field that acts as target for proton-photon interactions, can produce a neutrino flux peaking at $X*10^{-13}$ erg cm$^{-2}$ s$^{-1}$. But even in this particular case, these neutrino fluxes remain out of reach for the current neutrino observatories such as IceCube. This result is consistent with the non-detection of extreme blazars as point-like PeV neutrino emitters. The fact that the proton-synchrotron model is not associated with a significant neutrino emission is also in agreement with the theoretical results triggered by the recent detection of TXS 0506+056 as counterpart of the high-energy neutrino IC170922A \citep{icecube,gao18,keivani18,cerruti19}.

\section{\label{sec:discuss}Conclusions}
This paper reports the results of a multi-year observational campaign carried out by the MAGIC {C}ollaboration and aimed at a detailed {characterization} of the SEDs of ten EHBLs. The sources have been selected with different, complementary criteria and were observed with the MAGIC telescopes between 2010 and 2017. Observations of the archetypal EHBL 1ES~0229+200 between 2013 and 2017 were also included and used for comparison. Due to their relevance for the SED characterization in EHBLs, large part of the MAGIC data have been complemented by simultaneous {\it Swift}-XRT observations.  

The analysis of 262\,h of MAGIC data revealed a significant VHE gamma-ray signal from four sources: 1ES~1426+428, already detected by the HEGRA and VERITAS arrays, and the three new sources 1ES~2037+521, RBS~0723, and TXS~0210+515. In addition, a hint of VHE gamma-ray signal was found from RGB~J2042+244. The intrinsic (EBL-corrected) spectra are on average quite hard, indication of an extreme location of the second SED peak, exceeding the 100\,GeV range. The faint gamma-ray fluxes prevented a detailed time-resolved analysis. Since the SED peaks are shifted towards high energies, EHBLs are by definition faint and usually hard {\it Fermi}-LAT sources. Except for RBS~0921, from which only a hint of HE gamma-ray signal has been observed by {\it Fermi}-LAT, the spectral indices determined in time intervals centered on MAGIC observations range from $1.4 \pm 0.2$ to $1.9 \pm 0.2$.  Once corrected for the EBL absorption, the spectral indices of the VHE gamma-ray spectra range from $1.6 \pm 0.3$ to $2.7 \pm 1.2$. This  suggests a hard-TeV nature for all  detected sources but RBS~0723 whose spectrum is anyhow affected by large error bars and yet in agreement with the hard-TeV nature hypothesis. Among the new TeV-detected sources, TXS~0210+515 is the source with the hardest spectral index, making it a good target for deep exposure observations.

In the soft X-ray band, the analysis of all the available {\it Swift}-XRT data, including archival data, suggested only a limited variability, within a factor of two. The X-ray spectral indices anti-correlate with the flux levels, in agreement with a harder-when-brighter {behavior} typical for other TeV BL Lacs. 

For two sources (TXS~0210+515 and RGB~J2313+147) also the available {\it NuSTAR} data were analyzed, while {\it NuSTAR} data of 1ES~0229+200 covering MAGIC data window were adopted from literature. With its 3.0--79 keV energy coverage, {\it NuSTAR} is the ideal instrument to study and {characterize} EHBLs, even better if the data are analyzed in conjunction with {\it Swift}-XRT data allowing us to have a simultaneous fit of the X-ray spectrum from 0.5 to 79 keV (see Appendix B). In the case of TXS~0210+515, a clear evidence for a curved X-ray spectrum was found. The spectrum is well described by a log-parabola model, suggesting a position of the synchrotron peak at 7.1 $\pm$ 1.1\,keV. This confirms the extreme-synchrotron nature of the source, similar (but still less extreme) than 1ES~0229+200, for which a synchrotron peak at 9.1 $\pm$ 0.7\,keV has been estimated by \citet{2018MNRAS.477.4257C}. 
For RGB~J2313+147, the X-ray flux observed with {\it NuSTAR} is a factor of ten lower with respect to that of TXS~0210+515.
The joint XRT and \textit{NuSTAR} data are compatible with a power law spectrum with index larger than 2, and suggest a synchrotron peak located below 10$^{17}$\,Hz. This source was therefore very likely a standard HBL and not an EHBL during the observations.


All the SEDs were modeled with the single zone, conical-jet SSC model described by \citet[][and references therein]{asano18}. The six sources with spectral determination at VHE gamma rays, i.e. the four MAGIC detections, the hint-of-signal source, and the reference source 1ES~0229+200, were also modeled with two alternative scenarios: a leptonic scenario with a structured jet, the spine-layer model \citep{2005AA...432..401G}, and the proton synchrotron model described by \citet{cerruti15}. All the models provide a good description of the quasi-simultaneous multi-wavelength observational data. However, the resulting parameters differ substantially in the three scenarios. 

Main conclusion of the single-zone conical-jet SSC model applied to our data is that it requires a critically low magnetization, in tension with radio observations of nearby radio galaxies. The {spine-layer model} seems to provide a satisfactory solution to the magnetization problem, resulting in a quasi-equipartition of the magnetic field and matter in the emission zone. The proton-synchrotron model, instead, while still providing a good fit to the multi-wavelength data, results in  a highly magnetized jet, still far from equipartition. Therefore, with the current data set we cannot favour or disfavour any model considered.

Future observations of the EHBLs presented in this work (and of other EHBLs) will be essential for testing the emission models. Probing fast variability at VHE and  variability at different frequencies, in particular between the X-ray and VHE bands, is likely the most powerful tool at our disposal to test emission models. But given the faint signal at VHE with respect to the spectral capabilities of the current generation of IACTs, this is mostly a target for telescopes of future generations. In the meatime, coordinated multi-frequency monitoring and discovery of new VHE emitters belonging to the EHBL class is essential to prepare the ground for future discoveries.

\section*{Acknowledgements}

%
%
The authors are grateful to Luigi Costamante, for providing the data previously published.  Part of this work is based on archival data, software, or online services provided by the Space Science Data Center – ASI. This research has made use of data and/or software provided by the High Energy Astrophysics Science Archive Research Center (HEASARC), which is a service of the Astrophysics Science Division at NASA/GSFC and the High Energy Astrophysics Division of the Smithsonian Astrophysical Observatory.

We would like to thank the Instituto de Astrof\'{\i}sica de Canarias for the excellent working conditions at the Observatorio del Roque de los Muchachos in La Palma. The financial support of the German BMBF and MPG, the Italian INFN and INAF, the Swiss National Fund SNF, the ERDF under the Spanish MINECO (FPA2015-69818-P, FPA2012-36668, FPA2015-68378-P, {FPA2017-82729-C6-2-R}, FPA2015-69210-C6-4-R, FPA2015-69210-C6-6-R, AYA2015-71042-P, AYA2016-76012-C3-1-P, ESP2015-71662-C2-2-P, FPA2017-90566-REDC), the Indian Department of Atomic Energy, the Japanese JSPS and MEXT, the Bulgarian Ministry of Education and Science, National RI Roadmap Project DO1-153/28.08.2018 and the Academy of Finland grant nr. 320045 is gratefully acknowledged. This work was also supported by the Spanish Centro de Excelencia ''Severo Ochoa'' SEV-2016-0588 and SEV-2015-0548, and Unidad de Excelencia ``Mar\'{\i}a de Maeztu'' MDM-2014-0369, by the Croatian Science Foundation (HrZZ) Project IP-2016-06-9782 and the University of Rijeka Project 13.12.1.3.02, by the DFG Collaborative Research Centers SFB823/C4 and SFB876/C3, the Polish National Research Centre grant UMO-2016/22/M/ST9/00382 and by the Brazilian MCTIC, CNPq and FAPERJ. 

E.P. has received funding from the European Union's Horizon2020 research and innovation programme under the Marie Sklodowska--Curie grant agreement no 664931.

M.C. has received financial support through the Postdoctoral Junior Leader Fellowship Programme from la Caixa Banking Foundation, grant  n. LCF/BQ/LI18/11630012.

\software{MARS \citep{mars13}, enrico \citep{2013arXiv1307.4534S}, nustardas (v1.7.1) (\url{https://heasarc.gsfc.nasa.gov/docs/nustar/analysis})}


\vspace{0.5cm} 


\appendix
\section{\label{app:magic}MAGIC data analysis details}

MAGIC \citep{magicperf_1:2015} is a system of two IACTs designed to collect the UV-optical Cherenkov light generated when a gamma ray enters the atmosphere producing a shower of superluminal, charged particles. The two telescopes are located on the Canary island of La Palma, at 2200\,m altitude. With their large reflective surface of 17\,m diameter each, the MAGIC telescopes are designed to reach an energy threshold as low as 50\,GeV when operated in  standard trigger mode. Above 220\,GeV the integral sensitivity for point-like sources  is ($0.66\pm0.03$)\% of the Crab Nebula flux in 50\,h of observations, assuming a Crab Nebula-like spectrum. The angular resolution at those energies is {below} 0.07$\degree$, while the energy resolution is 16\%. The performance of the instruments and the details on the data analysis are fully described in \citet{magicperf_2:2015}. 

The main parameters influencing the energy threshold of the analysis are the zenith angle of the observations, and the {background} light conditions during data taking. 
Medium and high zenith angle observations ({above} 35$\degree$, and {above} 50$\degree$, respectively) are {characterized} by an increased energy threshold, due to the passage of the particle showers through a larger layer of atmosphere, but also due to an increased sensitivity at the highest energies related to the enlarged effective area \citep{magicperf_2:2015}.

{An higher level of background light due to the presence of the Moon strongly affects the energy threshold of the analysis. However, the performance of the telescopes system remains unaffected as long as the intensity of the moonlight is not too high \citep{magicperf_moon:2017}.} The data were analyzed using the MAGIC analysis and reconstruction software (MARS) package~\citep{mars13} that was  adapted to stereoscopic observations~\citep{mars09}. To look for a significant VHE gamma-ray excess, the standard variable, named $\theta^2$, was used, which is defined as the squared angular distance of the reconstructed shower direction with respect to source location in the camera. The typical signature of VHE gamma rays is peaking at low $\theta^2$ values, i.e. in the so-called `On' region in the camera, over the normalized cosmic-ray background, which is estimated from three equivalent `Off' regions, located at $90^\circ$, $180^\circ$ and $270^\circ$ with respect to the reconstructed source position in the camera.

\begin{figure}[ht!]
\centering
\includegraphics[width=0.50\textwidth]{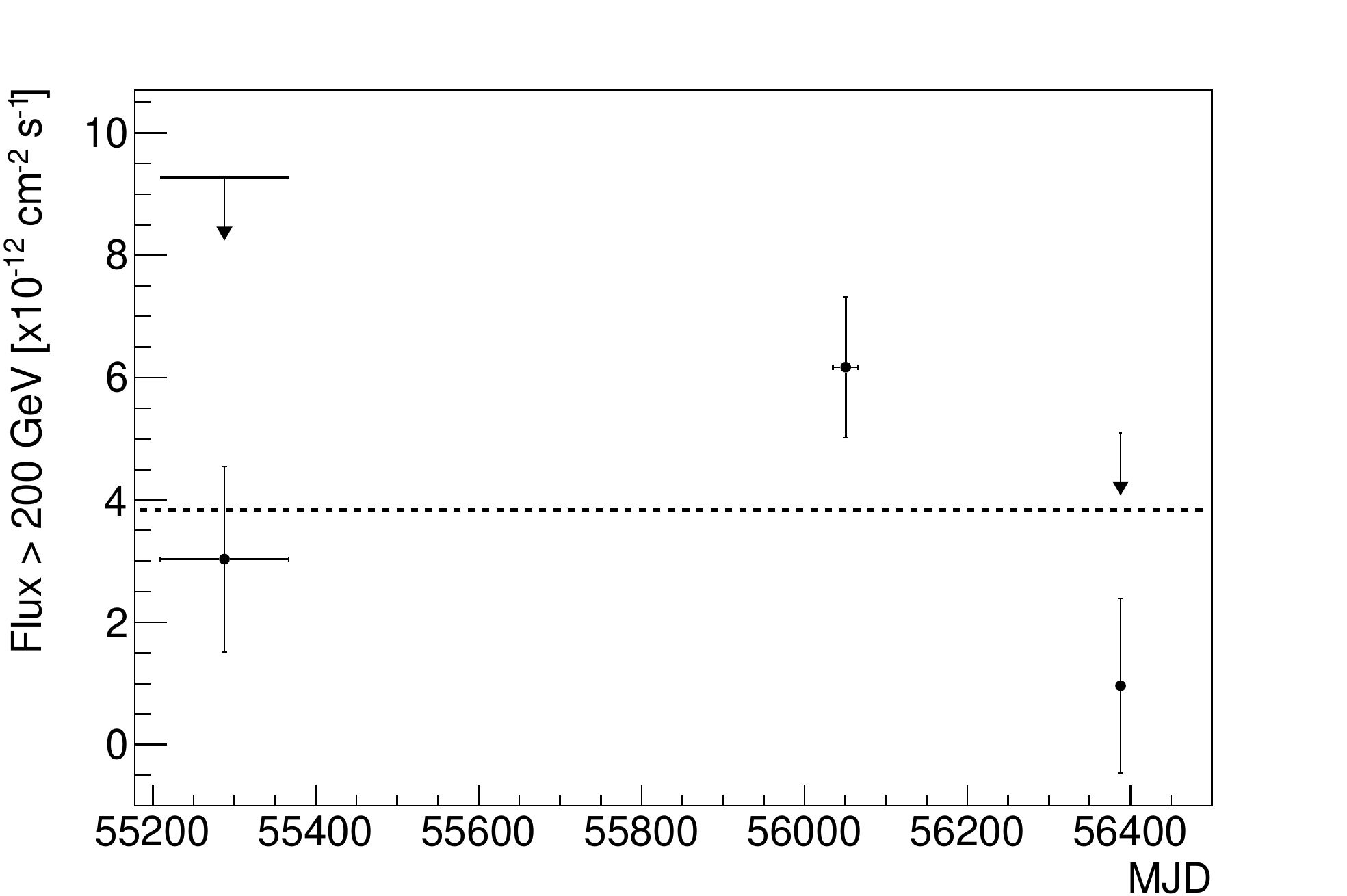}
\caption{\label{Fig:1426_LC}1ES~1426+428 multi-year MAGIC VHE gamma-ray light curve above 200 GeV between 2011 and 2013. Black downward-pointing arrows correspond to 95 per cent confidence upper limits, which were computed for the observations where the interval of the measured flux (black points) $\pm$ twice the error contains zero.}
\end{figure}

In Figure~\ref{Fig:1426_LC} the multi-year light curve of 1ES~1426+428 is displayed, reporting the average values measured from 2010, 2012, and 2013 observations. Only in 2012 the source was detected with a significance larger than 5 sigma, as reported in Table~\ref{tab:obs_table_magic}. The average flux above 200\,GeV is $(3.84\pm0.77)\times10^{-12}$\,cm$^{-2}$\,s$^{-1}$. From these data, the hypothesis of a constant flux cannot be excluded, especially if we take into account the systematic uncertainty on the integral flux.

A comparison of previous VHE gamma-ray observations based on the observed spectral indices, the differential fluxes at a given energy , and the integral fluxes for a common energy range was performed (see Table\,\ref{tab:spectr_comp_vhe}). The observed spectral indices are consistent within their statistical errors. The comparison on the differential and integral fluxes is based on the power-law fits to the observed differential spectra. The decorrelation energy of the spectral analysis of this work and an energy threshold of 200 GeV were used for the differential and integral flux calculations, respectively. The statistical errors reported for the flux calculations are overestimated as, for simplicity, the uncertainties of the fit parameters were considered to be uncorrelated when propagating the errors. On the other hand, the systematic uncertainties of the different instruments, which have been neglected in this comparison, might dominate over the statistical errors. It has also to be noted that the VHE gamma-ray spectra were determined for different energy ranges, which introduces a certain bias in the flux comparison. Given all these circumstances, a clear conclusion on the variability of the VHE gamma-ray flux of 1ES~1426+428 could not be drawn.

\begin{table*}[ht!]
\centering
\caption{\label{tab:spectr_comp_vhe} Comparison of the MAGIC results of 1ES 1426+428 from this work with previous VHE gamma-ray observations of this source.  {Columns from \textit{left} to \textit{right}}: year(s) of observation, energy range, fit parameters of the observed spectra, i.e., the flux normalization and the spectral index, differential flux and integral flux derived from the simple power-law fits. Only the statistical errors are reported.}
\begin{tabular}{ccccccc} 
\hline
\multirow{2}{*}{Year}    & Energy range &  $F_0$& \multirow{2}{*}{$\Gamma_{\mathrm{obs}}$} & $F_{E= 242{\rm{\,GeV}}}$ &$F_{(200 < E < 5000){\rm{\,GeV}}}$ & \\
              &[GeV]  &$\times10^{-12}$[cm$^{-2}$ s$^{-1}$ TeV$^{-1}$] & &$\times10^{-12}$[cm$^{-2}$ s$^{-1}$ TeV$^{-1}$] &$\times10^{-12}$[cm$^{-2}$ s$^{-1}$] \\
\hline
1998-2000$^1$    & $250-1000$ &$67 \pm 13_{E=400\rm{\,GeV}}$  & $3.6 \pm 0.6$ & $411 \pm 141$ & $63 \pm 16$\\
1999-2000$^2$ & $700-10000$ &$2.0 \pm 1.3_{E=1000\rm{\,GeV}}$  & $2.6 \pm 0.6$ & $80 \pm 86$& $16\pm 14$\\
2001$^3$ & $250-1700$ &$4.9 \pm 1.4_{E=1000\rm{\,GeV}}$  & $3.5 \pm 0.4$ & $703\pm 352 $ &$110 \pm 47$ \\
2012$^4$    & $100-1700$ & $25.6 \pm 0.1_{E=242\rm{\,GeV}} $ &  $2.6\pm 0.3$ &$25.6 \pm 0.1 $ &$5.2\pm0.7$ \\
\hline
\end{tabular}
\begin{tablenotes}
\item {1: \citet{cat00}; 2: \citet{hegra02}; 3: \citet{veritas01}; 4: this work; the integral flux reported here is calculated from the spectral fit and thus shows little variation from the averaged flux observed in 2012 (see Fig.~\ref{Fig:1426_LC}), whose calculation is based on the number of gamma-like excess events instead. However, within the statistical errors both values are consistent.}
\end{tablenotes}
\end{table*}

\section{\label{app:fermi_analysis}{\it Fermi}-LAT data analysis details}
The {\em Fermi} Large Area Telescope (LAT) \citep{2009ApJ...697.1071A} is a pair conversion telescope consisting of a $4\times4$ array of silicon strip trackers and tungsten converters and a Cesium Iodine (CsI) based calorimeter. The instrument is fully covered by a segmented anti-coincidence shield which provides a highly efficient vetoing against charged particle background events. The LAT is sensitive to gamma rays from $20\,{\rm MeV}$ to more than $300\,{\rm GeV}$ and normally operates in survey mode, covering the whole sky every three hours and providing an instantaneous field of view (FOV) of $2.4\,{\rm sr}$.

The LAT data were extracted from the weekly LAT data files available in the FSSC data center\footnote{\protect\url{https://fermi.gsfc.nasa.gov/ssc/data/access/}}.  For each data sample, only Pass 8 source-class photons detected within $15^\circ$ of the nominal position of the analyzed source were considered. Only events whose reconstructed energy lies between $1\,{\rm GeV}$ and $300\,{\rm GeV}$ were selected. The relatively high energy threshold was set to simplify the analysis of the two fields and remove contamination from secondary sources. This was particularly important in the case of  1ES~2037+502  due to its proximity to the galactic plane. Following the event selection recommendations from Cicerone\footnote{\protect\url{https://fermi.gsfc.nasa.gov/ssc/data/analysis/documentation/Cicerone/}}, only good data ({\tt (DATA\_QUAL>0)\&\&(LAT\_CONFIG==1)}) with zenith distance lower than $90^\circ$ were included.

For each data sample, the data were reduced and analysed using the open-source software package {\tt enrico} \citep{2013arXiv1307.4534S} as a wrapper for the {\em Fermi} {\tt ScienceTools} (version v10r0p5)\footnote{\protect\url{http://fermi.gsfc.nasa.gov/ssc/data/analysis/scitools}}. 
A summed binned likelihood analysis approach was followed splitting in PSF event types (0, 1, 2 and 3) with 10 bins per energy decade and using the instrument response functions (IRFs) {P8R2\_SOURCE\_V6}. All the 3FGL sources within  the region of interest (ROI) are included in the model, along with Galactic and isotropic models using {\tt gll\_iem\_v06.fits} and {\tt iso\_P8R2\_SOURCE\_V6\_v06.txt} files, respectively. The spectra of the sources were selected such to maximize the value of the likelihood while being physically sound, following the same method described in \citet{2019MNRAS.486.4233A}. All sources were modeled with attenuated spectral shapes using the EBL template from \cite{2008AA...487..837F}. For each analysis, the spectral parameters of all sources that are significantly detected within a radius of $3^\circ$ around the source of interest were left free in the fit in order to account for their possible variability. The parameters of the rest of the sources are fixed to the published 3FGL values. The {normalization} of the diffuse components was left free. 

\begin{table*}[ht!]
	\centering
		\caption{\label{tab:src_lat}Main spectral parameters from {3FGL}, 2FHL, and 3FHL {catalogs}.   {Columns from \textit{left} to \textit{right}: source name; HE gamma-ray flux in range of 1-100 GeV, spectral index for the power-law fit in range of 100 MeV - 100 GeV, and the detection significance reported by \citet[3FGL,][]{Acero15}; spectral index for the power-law fit $>10$ GeV reported by \citet[2FHL,][]{Ackermann16}; spectral index for the power-law fit $>50$ GeV reported by \citet[3FHL,][]{ajello18}.}}
 		
 		\begin{tabular}{lcccccc} 
 			\hline
 			\multirow{2}{*}{Source} & Flux$_{\mathrm{3FGL}}$ (1-100\,GeV) &\multirow{2}{*}{$\Gamma _{\mathrm{3FGL}}$} & Significance       &  $\Gamma _{\mathrm{3FHL}}$   & $\Gamma _{\mathrm{2FHL}}$   \\
             &  $\times 10^{-10}$ [photon\,cm$^{-2}$\,s$^{-1}$] &  & [$\sigma$] & $E > 10$\,GeV       & $E> 50$\,GeV       \\
 			\hline
            TXS~0210+515   & $4.17\pm0.85$ & $2.04\pm0.17$ & 7.2   & $1.55\pm0.22$ &$1.85\pm0.47$   \\
            TXS~0637-128 & $3.34\pm0.93$ & $1.51\pm0.16$& 8.0  & & $1.63\pm0.43$ \\
            BZB~J0809+3455 & $3.21\pm0.68$ & $1.67\pm0.13$ & 8.0  & $1.71\pm0.27$ & $1.09\pm0.61$  \\
            RBS~0723       & $5.06\pm0.85$ & $1.74\pm0.11$ & 10.5 &  $1.86\pm0.21$&$3.60\pm0.27$   \\
            1ES~0927+500   & $1.83\pm0.68$ & $1.45\pm0.21$ & 5.1  &  $1.97\pm0.32$& N.A.          \\
            RBS~0921       & N.A.  &     N.A.          &   N.A.   &  N.A.         & N.A.          \\   
            1ES~1426+428   & $6.60\pm0.84$ & $1.57\pm0.08$ & 16.7 &  $1.91\pm0.14$&$3.34\pm0.58$ \\
            1ES~2037+521   & $3.93\pm0.13$ & $1.89\pm0.21$ & 5.2  &  N.A.         &N.A.           \\
            RGB~J2042+244  & $5.15\pm0.96$ & $1.87\pm0.14$ & 8.4  &  $1.88\pm0.25$&N.A.      \\
            RGB~J2313+147  & $6.21\pm0.97$ & $1.76\pm0.11$ & 11.7 &  $1.57\pm0.43$&$3.56\pm1.31$ \\
            \hline
             1ES~0229+200 & $4.39\pm0.90$ & $2.02\pm0.15$ & 7.2 & N.A. &  N.A.\\
 			\hline
	 	\end{tabular}
\end{table*}

\begin{figure}[ht!]
\centering
\includegraphics[width=0.5\textwidth]{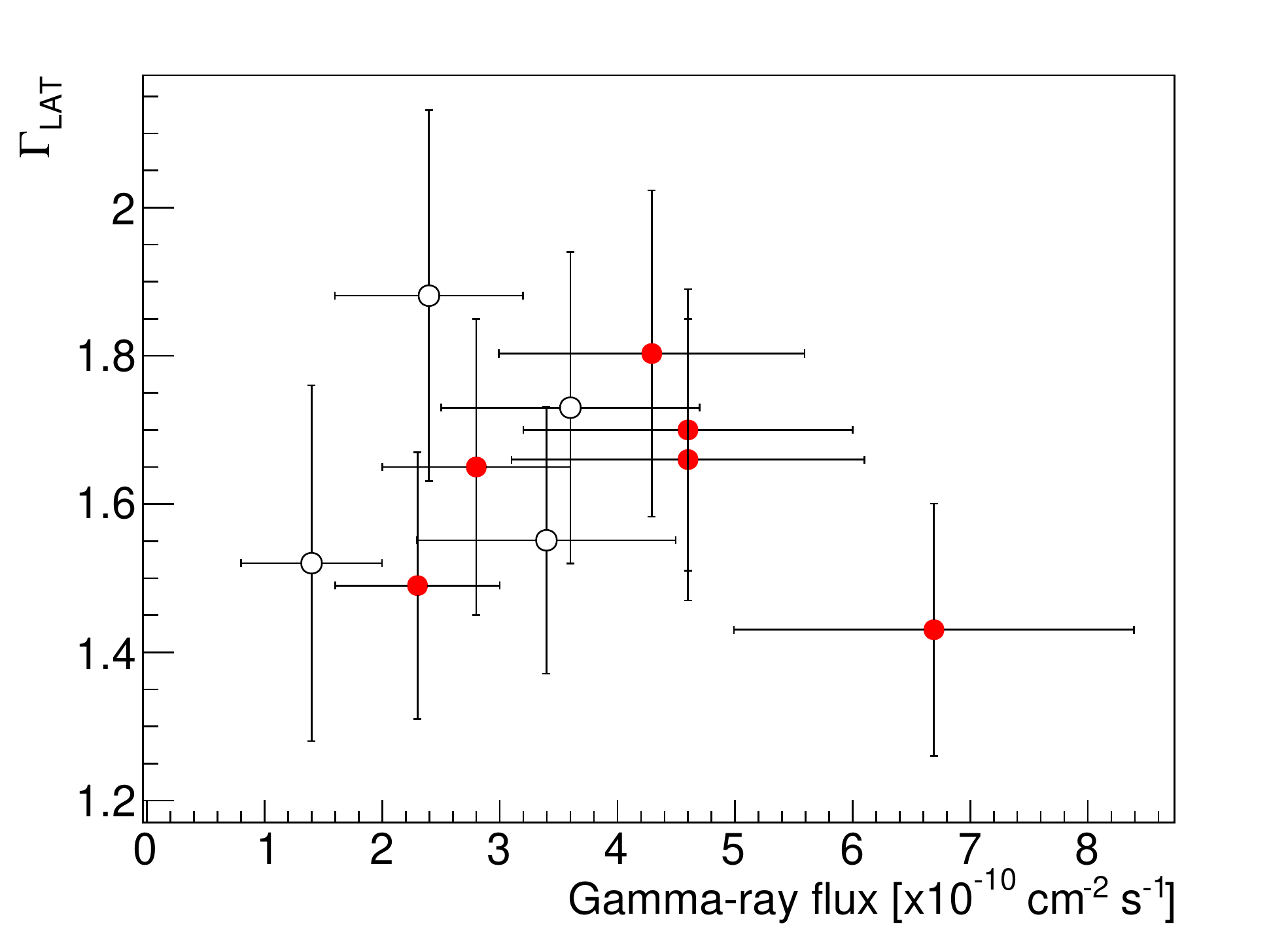}
\caption{\label{Fig:Fermi_correlation}Correlation study between the spectral index describing the average spectrum in the HE gamma-ray band and the integral HE gamma-ray flux between 1 and 100 GeV, both reported in Table~\ref{tab:src_lat}. Filled and open circles refer to sources detected and not detected at VHE gamma rays, respectively. The hint-of-signal source is considered here among the detected sources.}
\end{figure}

In order to investigate the relation of the \textit{Fermi}-LAT spectral properties on the extremeness at VHE we have compared the LAT spectral index and flux reported in Table~\ref{tab:spectra_LAT} for MAGIC detected and undetected sources (a similar study for the X-ray band is reported above). Our data, displayed in Figure~\ref{Fig:Fermi_correlation}, show that while the LAT spectral index does not have any effect on the detection probability, the flux seems to have a role: of the five detected sources, three were the brightest in the GeV band. This is quite regular and does not constitute a valid criterion for hard-TeV source selection (if the source is bright in LAT it is more likely to detect it also in the VHE range).
Interestingly, 1ES~0229+200 is instead the second faintest source of the sample in the GeV range. 

\section{\label{app:xrt_analysis}{\it Swift}-XRT data analysis details}

For the {\it Swift}-XRT data analysis of each source, the multi-epoch event list obtained by the XRT were downloaded from the publicly available SWIFTXRLOG ({\it Swift}-XRT Instrument Log\footnote{\url{https://heasarc.gsfc.nasa.gov/W3Browse/all/swiftxrlog.html}}) for both photon counting (PC) and window timing (WT) modes. The standard \textit{Swift}-XRT analysis procedure is described by \citet{2009MNRAS.397.1177E}. The PC data were processed using the procedure described by \citet{2017AA...608A..68F}. For the WT observation data, a box with length of 40 pixels at the centre of the source and aligned to the telescope roll angle was defined for the source region. The background region is defined by a box with length of 40 pixel aligned to the telescope roll angle and 100 pixel away from the centre of the source. 

There are open issues for analyzing the XRT data\footnote{\url{http://www.swift.ac.uk/analysis/xrt/digest_cal.php}}$^,$\footnote{\url{http://www.swift.ac.uk/analysis/xrt/rmfs.php}}, such as faults in the Silicon crystalline structure of the \textit{Swift}-XRT CCD. These open issues mostly affect the data obtained with WT mode. However, some of them (Charge Traps) still can affect the spectra observed during PC mode. In order to address these issues systematically in our data analysis, for both modes of observation using the $\chi^2$ likelihood method, the spectra of each observation were fitted assuming all possible combination of pixel-clipping patterns (XRT Grades) and point-spread-function (XRT response matrix files) \citep{2005SSRv..120..165B}. Simultaneously, two mathematical models \citep[i.e. power law and log parabola,][]{2004AA...413..489M} and fixed equivalent Galactic hydrogen column density reported by \citet{2005AA...440..775K} are assumed during spectral fitting procedure. The spectra are fitted in range of 0.3--10 keV except for the spectra of 1ES~2037+521. For 1ES~2037+521, the spectra is heavily absorbed by Galactic extinction at energies below 1.5\,keV. The current method, which is used in data analysis chain, can not address this issue correctly \citep{2013MNRAS.431..394W}. Therefore, we used 1.5--10\, keV as the energy range of spectral fitting for 1ES~2037+521. In total for each observation 6 and 16 spectra (PC and WT modes accordingly) are compared to each other and the best fitted model which describes the observation data is selected. Equivalent Galactic hydrogen column density of the sources are presented in Table~\ref{tab:src_table}. 

\begin{figure}
\centering
\includegraphics[width=1\textwidth]{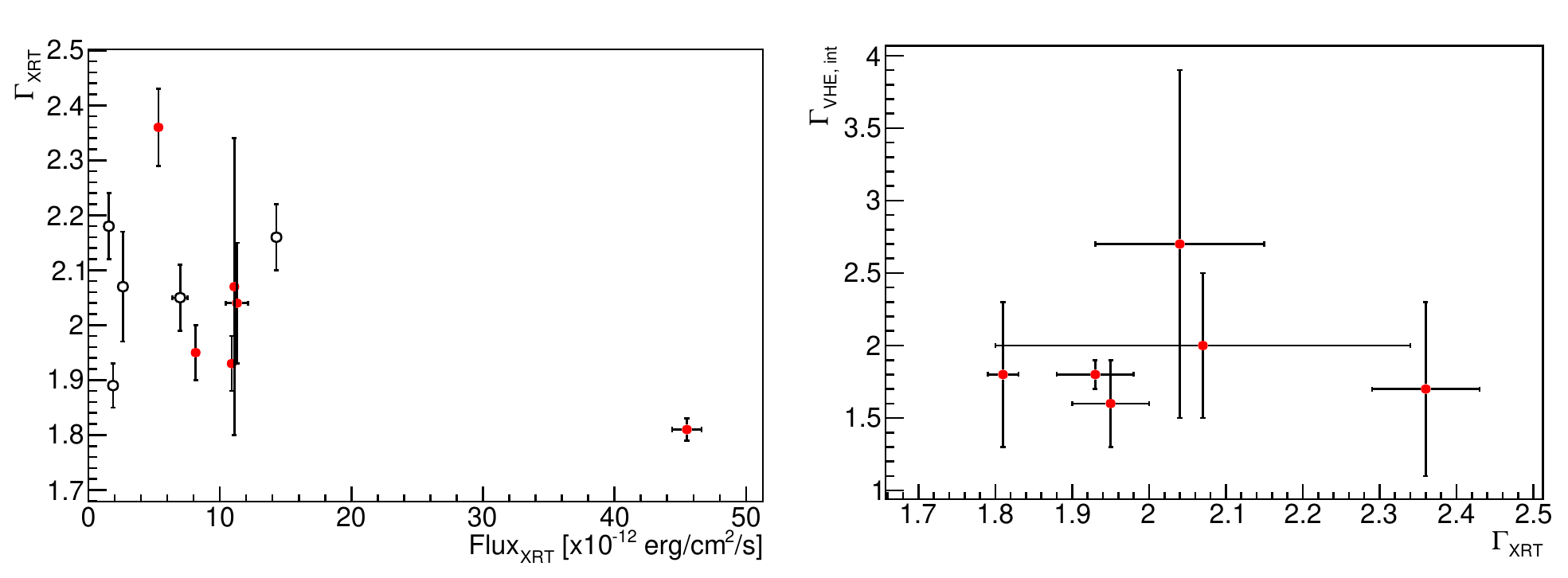}
\caption{\label{Fig:XRT_correlations}\textit{Left}: X-ray power law index versus integral flux (2-10\,keV) from the analysis of the average {\it Swift}-XRT spectra during MAGIC data taking, as reported in Table~\ref{tab:spectra_xrt}. Filled red markers are sources detected at VHE, while open markers represent sources still undetected at VHE. {The hint-of-signal source is considered here among the detected sources.} \textit{Right}: power law index in the X-ray band, from Table\,\ref{tab:spectra_xrt}, and  the power law index of the  the EBL-corrected (intrinsic) spectrum measured in the VHE gamma-ray band, reported in Table~\ref{tab:spectr_table_magic}.}
\end{figure}

Table~\ref{tab:obs_table_swift2} reports an example of the results obtained from {\it Swift}-XRT data. Small part of this sample was available in the public database, while the large majority of the observations was requested via ToO by the MAGIC team that performed quasi-simultaneous pointings with MAGIC telescopes. It is notable that the fit statistics are poor (i.e. $1.9<$reduced-$\chi^2<2.0$) for few of the observations (e.g. Tab.~\ref{tab:obs_table_swift2}, OBS ID 00046559005) due to the bad quality of raw data. In Table~\ref{tab:spectra_xrt}, the results of combining all the {\it Swift}-XRT data during MAGIC observation window for each are shown. The left panel of Figure~\ref{Fig:XRT_correlations} illustrates the distribution of X-ray photon index obtained from combined data sets in index-flux plane for the sources detected (Solid circle) and non-detected (open circles) in VHE gamma rays. No clear relation between the flux or the index and the detection probability at VHE gamma rays is evident (Fig.~\ref{Fig:XRT_correlations}, right panel).

\begin{table*}
\caption{\label{tab:obs_table_swift2}Example of the {\it Swift}-XRT results for RGB~J2042+244. Columns from \textit{left} to \textit{right}: day, observation ID, exposure time, spectral index of power law-model, $\chi^2$/d.o.f. of the fitted power-law model, spectral index of log-parabola model, curvature parameter of the fitted log-parabola model, $\chi^2$/d.o.f. of the log-parabola model,  Null-hypotheses of  probability of F-test, X-ray flux in range of 2-10 keV, and X-ray flux in range of 0.3-10 keV.}
\centering
\setlength{\tabcolsep}{0.37em}
\begin{tabular}{c|c|c|cc|ccc|c|cc} 
\hline
\multirow{2}{*}{Day}    & \multirow{3}{*}{OBS ID}  &  \multirow{2}{*}{Exposure}  & \multicolumn{2}{c|}{Power law}   & \multicolumn{3}{c|}{Log-parabola}      & \multirow{2}{*}{Prob.$^\star$}  & \multicolumn{2}{c}{Flux}   \\ 
\cline{4-5}
\cline{6-8}
\cline{10-11}

 &  &  & \multirow{2}{*}{$\Gamma$} & \multirow{2}{*}{$\chi ^2$/d.o.f.}   & \multirow{2}{*}{$\Gamma$} & \multirow{2}{*}{$\beta$} & \multirow{2}{*}{$\chi ^2$/d.o.f.}  &  &   F$_{2-10\mathrm{\,keV}}$    &   F$_{0.3-10\mathrm{\,keV}}$    \\ 
\cline{10-11}                        
[MJD]  &  &  [s]     &   &      &   &   &     & \%  &  \multicolumn{2}{c}{[$\times10^{-12}$ erg cm$^{-2}$s$^{-1}$]}         \\
\hline                                 
55939.73  &  00046559001  &  3024  & $ 2.05 \pm 0.07 $ &  28.2/28  &  &      &    &  0.7  & $ 4.7 \pm 0.5 $ & $ 8.4 \pm 0.5 $       \\
56299.56  &  00046559002  &  1176  & $ 1.75 \pm 0.18 $ &  5.2/4  &  &      &    &  47.0  & $ 4.1 \pm 1.1 $ & $ 6.1 \pm 1.0 $       \\
57192.04  &  00046559003  &  1985  & $ 1.93 \pm 0.07 $ &  29.5/27  &  &      &    &  99.6  & $ 9.2 \pm 0.8 $ & $ 15.1 \pm 0.9 $       \\
57194.08  &  00046559004  &  1641  & $ 1.95 \pm 0.08 $ &  17.5/22  &  &      &    &  67.1  & $ 8.6 \pm 0.8 $ & $ 14.3 \pm 1.0 $       \\
57196.08  &  00046559005  &  1791  & $ 1.87 \pm 0.07 $ &  50.9/26  &  &      &    &  19.1  & $ 9.8 \pm 1.0 $ & $ 15.6 \pm 1.0 $      \\
57968.03  &  00046559006  &  1558  &  &    & $ 1.85 \pm 0.12 $ & $ 0.98 \pm 0.30 $ &  15.2/20  &  0.1  & $ 5.1 \pm 0.7 $ & $ 12.0 \pm 0.9 $  \\
57979.33  &  00046559008  &  767  & $ 2.2 \pm 0.24 $ &  2.9/3  &  &  &    &  66.2  & $ 6.7 \pm 2.4 $ & $ 13.8 \pm 3.0 $       \\
57980.05  &  00046559009  &  1391  & $ 1.87 \pm 0.07 $ &  35.4/25  &  &  &    &  36.7  & $ 12.4 \pm 1.1 $ & $ 19.7 \pm 1.1 $       \\
57994.68  &  00046559010  &  787  & $ 2.03 \pm 0.09 $ &  21.0/17  &  &  &    &  47.5  & $ 11.3 \pm 1.4 $ & $ 19.7 \pm 1.4 $      \\
57996.94  &  00046559011  &  1371  & $ 2.01 \pm 0.08 $ &  32.1/21  &  &  &    &  3.1  & $ 8.1 \pm 0.8 $ & $ 14.0 \pm 0.9 $      \\
\hline
\end{tabular}
\begin{tablenotes}
\item $\star$ The log-parabola model is preferred over power-law model with $3\sigma$ confidence level if the F-test probability value is $<0.27\%$.
\tablecomments{Table~\ref{tab:obs_table_swift2} is published in its entirety in the machine-readable format.
      A portion is shown here for guidance regarding its form and content.}
\end{tablenotes}

\end{table*}

\begin{table*}
	\centering
		\caption{\label{tab:spectra_xrt}Main spectral parameters resulting from combining all {\it Swift}-XRT pointings during MAGIC observation window.  {Columns from \textit{left} to \textit{right}: source name, time interval of observation(s), X-ray flux in range of 2-10 keV, spectral index, curvature parameter of the log parabola model, fit statistics, and equivalent spectral index when log-parabola is the best-fit model.}}
 		
 		 \begin{tabular}{lcccccc} 
 			\hline
 			\multirow{2}{*}{Source} & Interval & F$_{(2-10keV)}$ & \multirow{2}{*}{$\Gamma$} & \multirow{2}{*}{$\beta$} & \multirow{2}{*}{$\chi^2$/d.o.f.} & \multirow{2}{*}{$\Gamma_{\mathrm{Equi}}$} \\
                   & [MJD] & [$\times$10$^{-12}$ erg cm$^{-2}$s$^{-1}$]& & & & \\
 			\hline
             TXS~0210+515 & 57370--58042 & $8.16 \pm 0.13$ & $1.69 \pm 0.02$& $0.23 \pm 0.04$ & 397.6/377  &  $1.95\pm0.05$ \\ 
            TXS~0637-128 & 57775--58023 & $14.32 \pm 0.32$ & $1.71 \pm 0.03 $ & $0.40 \pm 0.05$& 351.6/320  &$2.16 \pm 0.06 $ \\ 
            BZB~J0809+3455 & 57012--57317 & $1.88 \pm 0.09$ & $1.89 \pm 0.04$ & & 89.6/79 \\
            RBS~0723 & 56629-56985 & $11.30 \pm 0.84$ & $1.63 \pm 0.04$ & $0.36 \pm 0.09$ & 89.7/101 &$2.04 \pm 0.11$ \\
             1ES~0927+500  & 55641--55649 & $6.98 \pm 0.58$ & $2.05 \pm 0.06$& & 47.6/36 \\
             RBS~0921   & 57404--57483 & $2.63 \pm 0.11$  & $1.68 \pm 0.04$ & $0.34  \pm 0.08 $ &137.9/122  &$2.07 \pm 0.10$  \\
             1ES~1426+428 & 56039--56065 & $ 45.49 \pm 1.12 $ & $1.81  \pm  0.02$ & & 262.4/260   \\
             1ES~2037+521$^\star$ & 57658--57672 & $11.13 \pm 0.03$ & $1.46 \pm 0.17$ & $0.54  \pm 0.18$ &183.8/205 &$2.07 \pm 0.27$ \\
             RGB~J2042+244 & 57194--58055 & $5.33 \pm 0.15$ & $2.01 \pm 0.03$ &  $0.31  \pm 0.06$ &210.9/219&$2.36 \pm 0.07$\\
             RGB~J2313+147 & 57172 & $1.56 \pm 0.13$ & $2.18 \pm 0.06$ & &30.5/32 \\
 			\hline
 			1ES~0229+200  & 56566--57752 & $10.93 \pm 0.25$ & $1.50 \pm 0.02$& $0.38  \pm 0.04 $ &320.8/327 &$1.93 \pm 0.05$\\
 			\hline
	 	\end{tabular}
	 	\begin{tablenotes}
         \item $\star$ The range for spectral analysis is 1.5--10 keV (see text for details).
        \end{tablenotes}
\end{table*}

\section{\label{app:nustar_analysis}{\it Nustar} data analysis details}
The level 1 data products were processed with the {\it NuSTAR} Data Analysis Software (\texttt{nustardas}) package (v1.7.1). Cleaned event files (level 2 data products) were produced and calibrated using standard filtering criteria with the \texttt{NUPIPELINE} task, version 20180312 of the calibration files available in the {\it NuSTAR} CALDB and the OPTIMIZED parameter for the exclusion of the South Atlantic Anomaly passages.

For both objects, the source spectra were extracted from the cleaned-event files using a circle of 25 pixel ($\sim 60$ arcsec) radius, while the background was extracted from two distinct nearby circular regions of a 30-pixel ($\sim 70$\,arcsec) radius. The  ancillary  response  files  were  generated with the \texttt{numkarf} task, applying corrections for the point-spread-function losses, exposure maps and vignetting. The spectra were rebinned with a minimum of 20 counts per energy bin to allow for $\chi^2$ spectral fitting.

\begin{figure}
\centering
\includegraphics[width=1\textwidth]{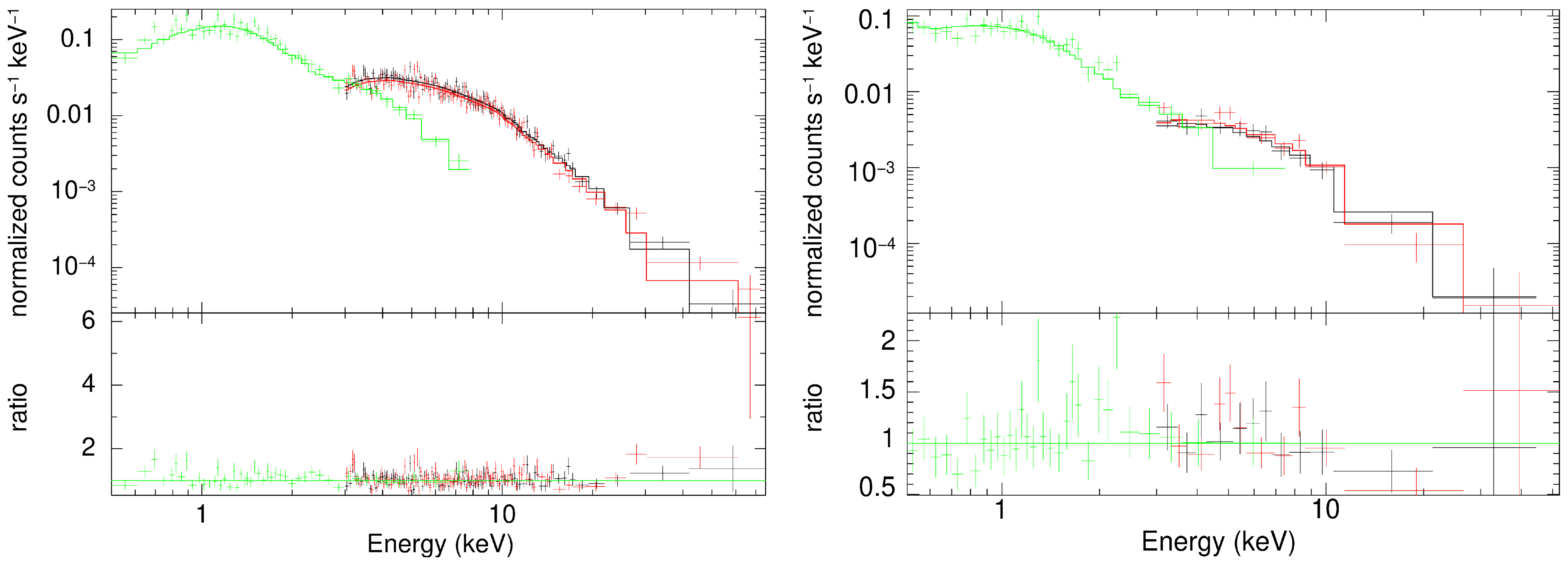}
\caption{\label{Fig:0210and2313_XRT_NuSTAR}Left: {\it NuSTAR} (black, FPMA, and red, FPMB, points) and {\it Swift}-XRT (green points) spectra and residuals for TXS~0210+515 collected on 2016 January 30 simultaneously fitted with a log parabola model.
Right: {\it NuSTAR} (red and black points) and {\it Swift}-XRT (green points) spectra and residuals of  RGB~2313+147 collected on 2015 May 30 simultaneously fitted with a power law model.}
\end{figure}

Table~\ref{Tab:nustar0210and2313} summarizes the results of the spectral analysis, described in the main text. The combined {\it NuSTAR} and {\it Swift}-XRT spectra are reported in Figure~\ref{Fig:0210and2313_XRT_NuSTAR}.
 
\begin{table}
\begin{center}
\caption{\label{Tab:nustar0210and2313}Summary of fits to the 0.5--79 keV Swift-XRT + NuSTAR spectrum of TXS 0210+515 and RGB~J2313+147. Fits also included absorption fixed at the Galactic value. Flux and synchrotron peak frequency are} given in units of erg\,cm$^{-2}$\,s$^{-1}$ and Hz, respectively.
\begin{tabular}[t]{llll}
\hline 
Model & Parameter &  TXS~0210+515 &  RGB~2313+147 \\
\hline
Power law & $\Gamma$ & $1.96 \pm 0.03$ & $2.32 \pm 0.10$  \\
      & Flux (0.5--79 keV)  &  $(2.17^{+0.03}_{-0.04}) \times 10^{-11}$   &  $(2.17^{+0.12}_{-0.16}) \times 10^{-12}$  \\
      &  $\chi^2$/\rm{d.o.f.}          &    387/311   &    62/63\\
\hline 
Log Parabola & $\alpha$ & $1.85 \pm 0.04$  & $2.37^{+0.11}_{0.10}$ \\
      & $\beta$ & $0.20 \pm 0.06$ & $0.35^{+0.18}_{-0.16}$ \\
      & $E_0$  & 3 keV (fixed) & 3 keV (fixed) \\
      & Flux (0.5--79 keV) &  $(1.92^{+0.06}_{-0.02}) \times 10^{-11}$ &  $(1.90^{+0.14}_{-0.11}) \times 10^{-12}$    \\
      &  $\chi^2$/\rm{d.o.f.}          &   348/310  &   50/62 \\
      &  $\log \nu_{synch}$      &     $18.24 \pm 0.07$           &  $17.33 \pm 0.16$  \\
\hline
\end{tabular}
\end{center}
\end{table}

\section{{Spectral energy distribution parameters}}

{In this section, the SED model parameters are reported in details for the three adapted scenarios. Only the parameters which are left to vary are reported here. The fixed parameters are described in Section~\ref{sec:modelling}. The results of the single zone, conical-jet model for all the sources of the study are listed in Table~\ref{param_tab}. Table~\ref{tab:spine} presents the parameters of the spine-layer model for the sources with a spectral determination, and Table~\ref{tab:hadronic-model} those of the proton-synchrotron model. In the former case, the solution is degenerated and a range is proposed for some of the parameters.}

\begin{table*}
\caption{\label{param_tab}Model parameters and obtained physical values for the SSC conical jet scenario. {Columns from \textit{left} to \textit{right}: source name; break and maximum electron Lorentz factor; spectral index of the electron energy distribution below and above $\gamma _{br}$; magnetic field; electron luminosity; bulk Lorentz factor of the jet; synchrotron and IC peak frequency resulting from the model; Compton dominance parameter; ratio between the magnetic and electron energy density evaluated at the radius where the electron injection shuts down; and the source detection status at VHE gamma rays (Y: detected, N: not detected, and H: hint of signal)}}
\centering
\begin{tabular}{lcccccccccccc}
\hline
\multirow{2}{*}{Source name} & $\gamma_{\rm br}$ & $\gamma_{\rm max}$  & \multirow{2}{*}{$p_1$} & \multirow{2}{*}{$p_2$} & $B_0$ & $L_{\rm e}\times10^{44}$   & \multirow{2}{*}{$\Gamma$} & $\log (\nu_{\rm{syn,pk}}) ^\star$ & $\log (\nu_{\rm{IC,pk}})^\star$ & \multirow{2}{*}{CD$^{\star \dagger}$} & $U_B$/$U_{\rm e}^\star$ & \multirow{2}{*}{VHE?} \\
			
 & [$\times 10^{5}$] &  [$\times10^{6}$]  &  &  & [G] & [$\mbox{erg}~\mbox{s}^{-1}$]   &  & [Hz] & [Hz] &  &  [$\times 10^{-3}$]  & \\
                            
\hline                            
TXS~0210+515   & 10.0 & 20.0  & 2.5 & 3.0 & 0.04 & 6.50 & 20 & 18.3 & 25.7 & 0.18 & 0.19  & Y \\
TXS~0637-128   & 5.0  & 20.0  & 1.8 & 3.0 & 0.25 & 0.80 & 10 & 17.7 & 25.4 & 0.2  & 81.00 & N \\
BZB~J0809+3455 & 1.0  & 3.0   & 1.8 & 3.0 & 0.04 & 0.89 & 10 & 16.4 & 25.4 & 0.84 & 1.40  & N \\
RBS~0723       & ---  & 2.0   & 2.2 & --- & 0.11 & 4.90 & 20 & 18.1 & 25.8 & 0.37 & 1.90  & Y \\
1ES~0927+500   & 3.0  & 3.0   & 1.5 & 2.5 & 0.13 & 0.71 & 10 & 17.6 & 25.9 & 0.25 & 23.00 & N \\
1ES~1426+428   & ---  & 2.0   & 2.0 & --- & 0.20 & 1.30 & 20 & 18.2 & 25.8 & 0.14 & 26.00 & Y \\
1ES~2037+521   & ---  & 2.0   & 2.1 & --- & 0.02 & 2.30 & 20 & 18.1 & 26.4 & 0.33 & 0.14  & Y \\
RGB~J2042+244  & ---  & 0.3   & 2.0 & --- & 0.07 & 1.80 & 20 & 17.1 & 25.6 & 0.36 & 2.30  & H \\
RGB~J2313+147  & 0.8  & 20.0  & 2.0 & 3.5 & 0.09 & 1.60 & 20 & 16.5 & 25.3 & 0.37 & 3.90  & N \\
\hline
1ES~0229+200   & 10.0 & 300.0 & 1.9 & 3.0 & 0.06 & 1.10 & 20 & 18.6 & 26.6 & 0.13 & 2.50  & Y \\
\hline
\end{tabular}
\begin{tablenotes}
\item $\star$ {These quantities are derived quantities, and not model parameters.}
\item $\dagger$ {The ratio of $\nu L_\nu$ at the IC peak to that at the synchrotron peak.}
\end{tablenotes}
\end{table*}

\begin{table*}
\centering
\caption{\label{tab:spine}Model  parameters  and  obtained  physical  values  for  the  spine-layer scenario  for the sources with VHE gamma rays spectral determination. Columns from \textit{left} to \textit{right}: source name; break and maximum electron Lorentz factor; spectral index of the electron energy distribution below and above $\gamma _b$; magnetic field; normalization of the electron distribution; radius of the emission zone; ratio between the magnetic and electron energy density of the layer; kinetic luminosity of the jet.}

\begin{tabular}{lcccccccccccc}
\hline
\multirow{2}{*}{Source name} & $\gamma _{\rm b}$ & $\gamma _{\rm max}$ & \multirow{2}{*}{$n_1$} & \multirow{2}{*}{$n_2$} & $B$ & $K$  &$R \times 10^{15}$ & \multirow{2}{*}{$U_B$/$U_{\rm e}^\star$} & $L_{\rm j}^\star \times10^{42} $ \\
 & [$\times 10^{4}$] & [$\times 10^{6}$] &  &  & [G] & [cm$^{-3}$] & [cm] &  & [erg s$^{-1}$] \\
\hline                   
TXS 0210+515  & 33.0 & 0.8 & 1.40 & 2.30 & 0.15 & 25.0 & 5.1 & 1.25 & 2.50  \\
RBS 723       & 0.3  & 0.8 & 1.40 & 2.30 & 0.35 & 15.0 & 5.1 & 1.17 & 14.60 \\
1ES 1426+428  & 3.0  & 2.0 & 1.40 & 2.90 & 0.34 & 3.5  & 7.1 & 1.07 & 20.50 \\
1ES 2037+521  & 13.0 & 2.0 & 1.40 & 3.00 & 0.40 & 2.9  & 1.3 & 0.75 & 0.97  \\
RGB J2042+244 & 2.0  & 2.0 & 1.40 & 2.95 & 0.30 & 3.0  & 4.8 & 1.21 & 7.00  \\
\hline
1ES 0229+200  & 13.0 & 6.0 & 1.40 & 3.40 & 0.40 & 2.6  & 3.2 & 0.74 & 6.30  \\
\hline
\end{tabular}
\begin{tablenotes}
\item $\star$ {These quantities are derived quantities, and not model parameters.}
\end{tablenotes}

\end{table*}

\begin{table*}
\centering
\caption{\label{tab:hadronic-model}Model  parameters  and  obtained  physical  values  for the hadronic scenario for the sources with VHE gamma rays spectral determination. Columns from \textit{left} to \textit{right}: minimum electron Lorentz factor; spectral index of the electron/proton energy distribution below and above $\gamma_{break}$; magnetic field; normalization of the electron distribution; radius of the emission zone; maximum proton Lorentz factor; efficiency of the acceleration mechanism ; magnetic energy density; ratio between the proton and magnetic energy density; luminosity of the emission region.}
\setlength{\tabcolsep}{0.295em}
\begin{tabular}{lccccccccccc}
\hline
\multirow{2}{*}{Source name} & $\gamma_{e,\text{max}}$ & \multirow{2}{*}{$\alpha_{1}$} & \multirow{2}{*}{$\alpha_{2}$} & $B$ & $K_e\times10^{-3}$ & $R\times10^{14}$ & $\gamma_{p,\textnormal{max}}$ & $\eta$ & $u_B^\star$ & $u_p$/$u_B^\star $ & $L^{\star \dagger} \times10^{46} $ \\
 & [$\times 10^4$] &  &  & [G] & [cm$^{-3}$] & [cm] & [$\times 10^9$] & [erg cm$^{-3}$] & [$\times 10^{-5}$] & [$\times 10^{-5}$] & [erg s$^{-1}$]  \\
\hline                      
TXS 0210+515  & 1.0--15.9 & 1.30 & 2.30 & 1.9--468 & 0.002--4890  & 1--1480 & 1.7--48.9 & 0.06--4.9   & 0.15--8710 & 0.008--47.8 & 0.10--48.1  \\
RBS 0723      & 1.1--16.5 & 1.25 & 2.25 & 2.1--468 & 0.035--68640 & 1--1300 & 1.6--28.0 & 0.12--3.5   & 0.18--8710 & 1.1--1300 & 0.10--32.4    \\
1ES 1426+428  & 1.2--15.9 & 1.25 & 2.25 & 2.0--344 & 0.09--120000 & 1--1380 & 1.6--21.0 & 0.07--1.7   & 0.17--4710 & 2.8--1070 & 0.11--18.2    \\
1ES 2037+521  & 1.1--15.6 & 1.30 & 2.30 & 2.0--401 & 0.002--7810  & 1--1480 & 1.6--29.2 & 0.16--6.6   & 0.15--6410 & 0.06--103 & 0.10--35.3    \\
RGB J2042+244 & 1.0--15.6 & 1.50 & 2.50 & 2.0--468 & 0.09--150000 & 1--1410 & 1.6--33.5 & 0.80--38.0  & 0.16-- 871 & 0.06--234 & 0.11--46.5    \\
\hline                      
1ES 0229+200  & 1.1--13.7 & 1.10 & 2.10 & 2.8--468 & 0.004--11130 & 1--1360 & 1.9--33.2 & 0.004--0.14 & 0.31--8710 & 0.11--140 & 0.15--45.6    \\
\hline
\end{tabular}
\begin{tablenotes}
\item $\star$ {These quantities are derived quantities, and not model parameters.}
\item$\dagger$ {The luminosity of the emitting region has been calculated as $L=2 \pi R^2c\Gamma_\textnormal{bulk}^2(u_B+u_e+u_p)$, where $\Gamma_\textnormal{bulk}=\delta/2$, and $u_B$, $u_e$, and $u_p$, the energy densities of the magnetic field, the electrons, and the protons, respectively.}
\end{tablenotes}
\end{table*}

\section{\label{app:SED0921}RBS~0921: spectral energy distribution}

The following Figure\,\ref{fig:0921_SED} show the broad band SED of RBS~0921. {Due to the lack of gamma-ray spectral data both in the HE and VHE bands, the SED of this source was not considered for modeling.}

\begin{figure}[ht!]
\centering\includegraphics[width=0.49\textwidth]{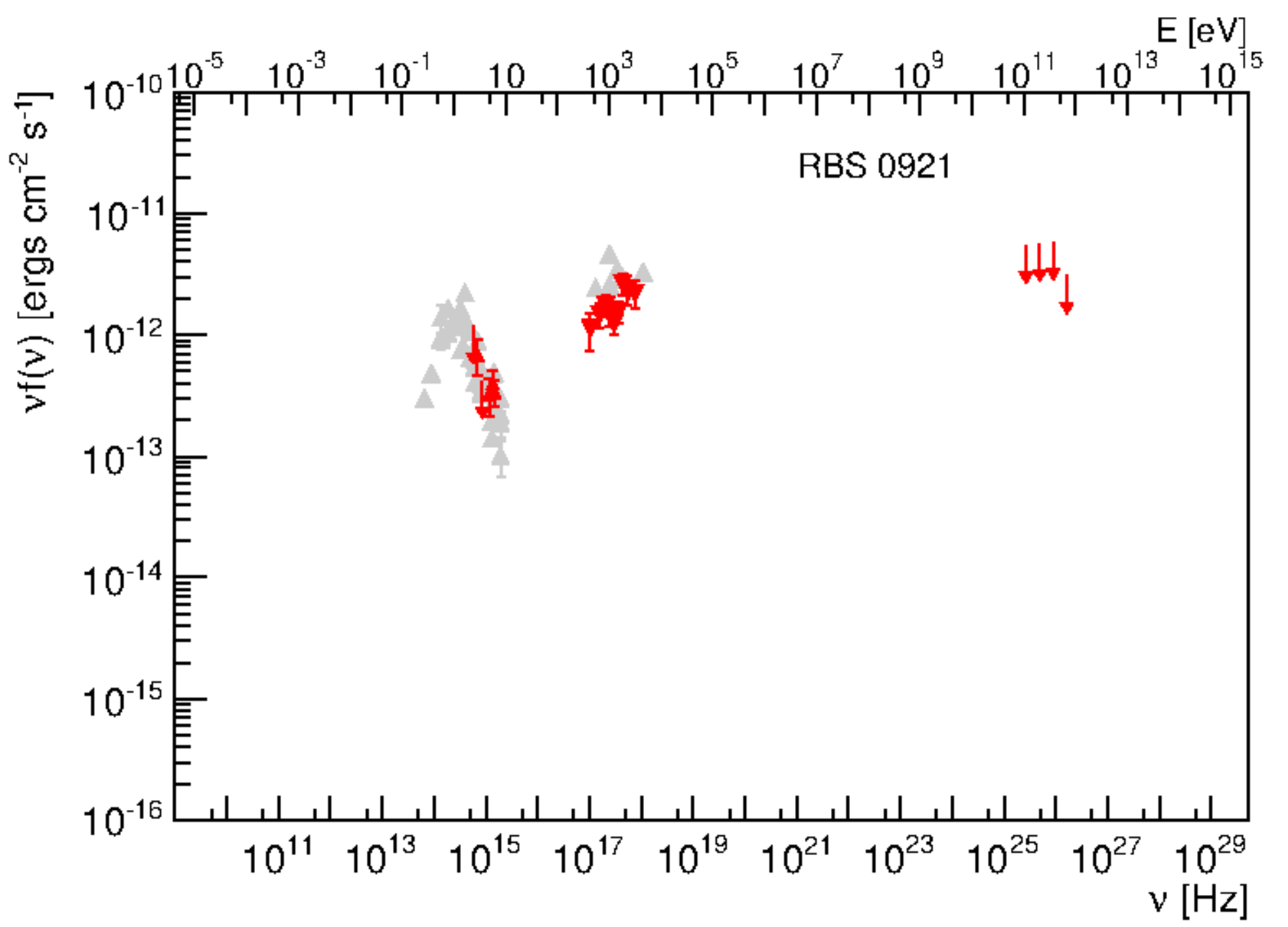}
\caption{\label{fig:0921_SED}SED of RBS~0921, the unique source of the sample still undetected at gamma rays.}
\end{figure}

\section{\label{app:neutrinoSED}Maximum neutrino flux expectations}

\begin{figure}[ht!]
\centering\includegraphics[width=0.60\textwidth]{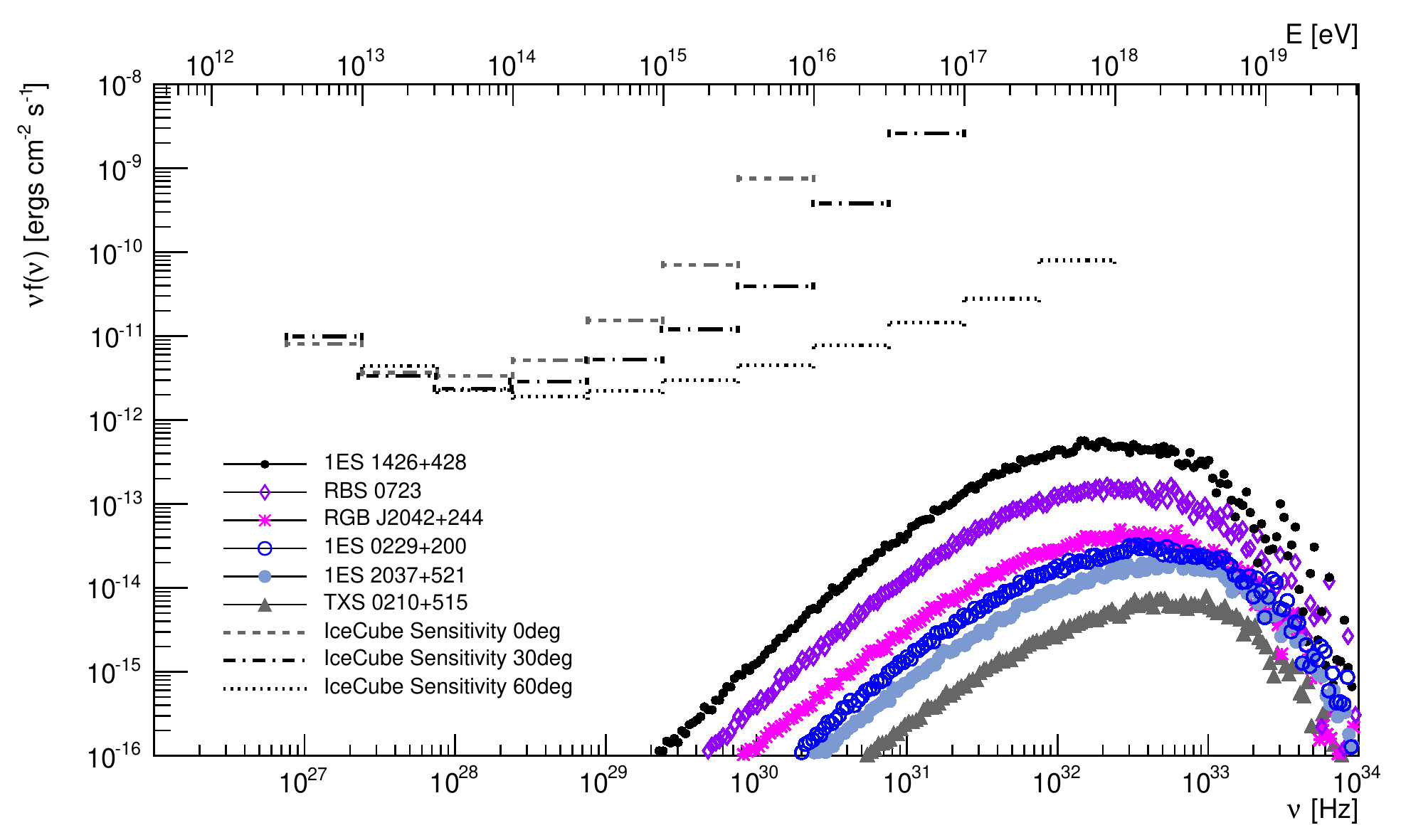}
\caption{\label{fig:MaxNeuFlux} The maximum neutrino flux expected from the proton-synchrotron model applied to six EHBLs. The IceCube sensitivity \citep{Aartsen2019} for three different declinations is also represented. }
\end{figure}

 In Figure\,\ref{fig:MaxNeuFlux} the  maximum  neutrino  flux expected from the proton-synchrotron model for the six sources considered is reported.

\label{lastpage}
\end{document}